\documentclass[twocolumn,prd,superscriptaddress,altaffilletter,nofootinbib]{revtex4-2}
\usepackage{float}
\usepackage{amsmath}
\usepackage{amssymb}
\usepackage{amsfonts}
\usepackage{comment}
\usepackage{graphicx}
\usepackage[colorlinks, pdfborder={0 0 0}, citecolor=BrickRed, linkcolor=BrickRed]{hyperref} 
\usepackage{interval}
\usepackage[nolist]{acronym}
\usepackage[usenames,dvipsnames]{xcolor}
\usepackage{xspace}
\usepackage{subfigure}
\usepackage{tikz}
\usepackage[utf8]{inputenc}
\usepackage[T1]{fontenc}
\usepackage{url}
\usepackage[squaren]{SIunits}
\usepackage{flushend}
\renewcommand{\footnoterule}{\vfill\kern -3pt \hrule width 0.4\columnwidth \kern 2.6pt}
\usetikzlibrary{arrows,shapes,trees,decorations.pathreplacing}
\usepackage{enumitem}
\usepackage{xcolor}

\newcommand{\aplus} {\texttt{A+}\xspace}
\newcommand{\voy} {\texttt{VK+Voy}\xspace}
\newcommand{\et} {\texttt{HLKI+E}\xspace}
\newcommand{\cefo} {\texttt{VKI+C}\xspace}
\newcommand{\EC} {\texttt{KI+EC}\xspace}
\newcommand{\ECS} {\texttt{ECS}\xspace}

\allowdisplaybreaks

\begin{document}

\title{Finding cosmic anisotropy \\ with networks of next-generation gravitational-wave detectors}


\author{Bryce~Cousins} 
\email[Corresponding author: ]{brycec2@illinois.edu}
\affiliation{Department of Physics \& Illinois Center for Advanced Studies of the Universe, University of Illinois Urbana-Champaign, Urbana, IL 61801, USA}
\affiliation{Institute for Gravitation and the Cosmos, The Pennsylvania State University, University Park, PA 16802, USA}

\author{Arnab~Dhani}
\affiliation{Institute for Gravitation and the Cosmos, The Pennsylvania State University, University Park, PA 16802, USA}
\affiliation{Department of Physics, The Pennsylvania State University, University Park, PA 16802, USA}
\affiliation{Max Planck Institute for Gravitational Physics (Albert Einstein Institute), Potsdam Science Park, Potsdam, Germany}

\author{Bangalore~S.~Sathyaprakash}
\affiliation{Institute for Gravitation and the Cosmos, The Pennsylvania State University, University Park, PA 16802, USA}
\affiliation{Department of Physics, The Pennsylvania State University, University Park, PA 16802, USA}

\author{Nicolás~Yunes} 
\affiliation{Department of Physics \& Illinois Center for Advanced Studies of the Universe, University of Illinois Urbana-Champaign, Urbana, IL 61801, USA}


\begin{abstract}
The standard cosmological model involves the assumption of isotropy and homogeneity, a principle that is generally well-motivated but is now in conflict with various anisotropies found using independent astrophysical probes. These anisotropies tend to take the form of dipoles; while some can be explained by simple kinematic effects, many others are not fully understood. Thus, generic phenomenological models are being considered, such as a dipole in the luminosity distance. We demonstrate how such a dipole could be measured using gravitational waves from binary neutron star mergers observed by six different networks of gravitational-wave detectors, ranging from upgraded LIGO detectors to anticipated next-generation ground-based observatories. We find that, for example, a network of three next-generation detectors would produce strong constraints on a dipole's amplitude ($\sim 13\%$) and location ($\sim 84$ deg$^2$) after just one year of observing. We demonstrate that the constraints scale with the number of detections, enabling projections for multiple years of observing. Our findings indicate that future observations of binary neutron star mergers would improve upon existing dipole constraints, provided that at least one next-generation detector is built. We also assess directional sensitivity of the dipole measurements by varying the dipole's location on a grid across the sky. We find that for a network of three next-generation detectors, the range of the constraints is only $\lesssim 1.2\%$ for the amplitude and $\lesssim 4\%$ for the location, indicating that the location of the dipole will not greatly impact our ability to measure its effects.
\end{abstract}

\maketitle

\section{Introduction} \label{sec:intro}

The standard $\Lambda$CDM cosmological model is based on the assumption of isotropy and homogeneity. This is known as the Cosmological Principle (CP) and has been supported by observations of the Cosmic Microwave Background (CMB)~\cite{planckcollaborationPlanck2018Results2020a, planckcollaborationPlanck2018Results2020}, which lends credence to the use of the Friedman-Lemaître-Robertson-Walker (FLRW) metric to describe the Universe. However, the CP's assumption of isotropy is in growing conflict with various anisotropies found using a range of astrophysical probes (reviewed, e.g., in Sec. VIII.F of \cite{abdallaCosmologyIntertwinedReview2022}, Sec. III.3 of \cite{perivolaropoulosChallengesLambdaCDM2022}, or Sec. IV of \cite{aluriObservableUniverseConsistent2023}). These anisotropies tend to have the multipolar structure of a dipole, suggestive of a preferred axis in the Universe. A cosmic dipole can generally be modeled as a dimensionless vector, possessing a direction (i.e.,~the axis along which the dipole's effects are most pronounced) and a dimensionless amplitude. This amplitude is generally taken to reflect a linear modulation of an observable relative to what is expected for the isotropic case. For example, if the distances or number counts of astrophysical objects have a $\pm 10\%$ deviation from the prediction of an isotropic cosmology along some axis, then the effect can be modeled phenomenologically as a dipole with an amplitude of $0.1$ with a direction along the observed axis.

Some dipoles can be understood purely by kinematics---such as the relative motion of the Earth---but the origins of other dipoles are not always fully understood. The confirmation of a non-kinematic dipole would not only violate the CP, but may also indicate the presence of new physics beyond the standard cosmological model. In light of this and the wide variety of observed dipole anisotropies, we first classify and discuss notable anisotropies (summarized in Table \ref{tab:dipoles}).

Kinematic anisotropies are the result of the relative motion between the source and observer frames, which, in turn, can be dominated by the motion of the source or observer.  An example of the former is the relative motion of a source due to its peculiar velocity or a larger-scale, ``bulk flow'' velocity; an example of the latter is Earth's motion relative to some frame (e.g., the CMB frame). In the most general case, both the source and the observer could be moving relative to some other reference frame. Regardless of its exact cause, a kinematic dipole leads to anisotropies in astrophysical observations that encode the direction and magnitude of motion between the frames.

A kinematic dipole would induce two effects on astrophysical observations~\cite{ellisExpectedAnisotropyRadio1984}. First, an angular aberration occurs along the axis of motion. This angular aberration distorts the angle at which a source is observed, concentrating more sources along the axis of forward motion. Second, a Doppler boost occurs relative to the axis of motion, which increases the observed intensity of sources in the hemisphere containing the direction of motion. Both effects result in a modified number of detections relative to the dipole axis, i.e., more detections in the direction of motion, and less in the direction opposite of motion. Quantifying a dipole by analyzing these effects thus allows the relative velocity to be measured, assuming that the dipole is entirely due to kinematics.

However, there are now observations of dipoles that are larger than what is expected from pure kinematics~\cite{bengalyProbingCosmologicalPrinciple2018,secrestTestCosmologicalPrinciple2021,horstmannInferenceCosmicRestframe2022}, suggesting that there may be anisotropies that deviate from the predictions of $\Lambda$CDM. This tension arises when attempting to interpret the dipoles observed in late-universe probes as arising from the same kinematics as the CMB kinematic dipole. Within $\Lambda$CDM, kinematic dipoles observed in both CMB and late-universe probes are expected to be generated in exactly the same way by Earth's orbital motion. For CMB observations, kinematics affect the measurement of the temperature by $\pm 0.123\%$, corresponding to a dipole of amplitude $1.23 \times 10^{-3}$, which matches theoretical expectations~\cite{planckcollaborationPlanck2018Results2020a}. When considering the same kinematics for radio galaxies as a late-universe probe, the number counts of galaxies should be modified by $\sim \pm 0.4\%$ along the direction of Earth's motion as computed in~\cite{bengalyProbingCosmologicalPrinciple2018}, corresponding to a dipole amplitude of $4 \times 10^{-3}$. However, as detailed in~\cite{bengalyProbingCosmologicalPrinciple2018}, the observed dipole amplitude is a factor of $\sim 17$ larger than expected ($7.0 \times 10^{-2}$), indicating that the kinematic interpretation may not entirely explain the observed dipole. A similar tension exists for quasars~\cite{secrestTestCosmologicalPrinciple2021} and supernovae~\cite{horstmannInferenceCosmicRestframe2022}. These tensions---which we denote as ``type K/I'' in Table~\ref{tab:dipoles}---indicate some unaccounted-for systematics or some intrinsic component of the matter-based dipoles that is not entirely due to kinematics.

\begin{table*}[!htp]
    \caption{Constraints on notable dipoles from recent studies. Each dipole is labeled as kinematic (K) if its amplitude is fully explained by Earth's motion relative to the CMB frame, as intrinsic (I) if kinematics do not affect the amplitude measurement, or as kinematic/intrinsic (K/I) if kinematics affect the amplitude measurement but do not fully explain it. Values of significance were reproduced from the original sources as either standard deviation or confidence level (C.L.). The row with a superscript star (*) was obtained using the \texttt{Commander} results from Table~25 and Fig.~34 in \cite{planckcollaborationPlanck2018Results2020}. Observe the presence of a large number of dipoles whose source is not yet fully understood.}
    \begin{tabular}{|c|c|c|c|c|c|c|c|}
    \hline
    dipole anisotropy & \multicolumn{1}{c|}{type}  & \multicolumn{1}{c|}{redshift} & \multicolumn{1}{c|}{amplitude} & \multicolumn{1}{c|}{$l \degree$}   & \multicolumn{1}{c|}{$b\degree$}   & \multicolumn{1}{c|}{significance} \\
    \hline
    \hline
    CMB solar dipole \cite{planckcollaborationPlanck2018Results2020a} & K & N/A & $1.23 \pm 0.00036 \times 10^{-3}$ & $264.02 \pm 0.01$ & $48.253 \pm 0.005$ & N/A \\
    CMB temperature asymmetry* \cite{planckcollaborationPlanck2018Results2020}  & I & N/A & $2.3^{+0.8}_{-0.4} \times 10^{-2}$ & $220 \pm 25$ & $-5 \pm 25$ & $\sim 2.7 \sigma$ \\ 
    SNe \cite{horstmannInferenceCosmicRestframe2022} (CMB frame) & K & $0.01 - 1$ & $8 \pm 1.7 \times 10^{-4}$ & $242 \pm 16$ & $59 \pm 19$ & $68\%$ C.L. \\
    SNe \cite{singalPeculiarMotionSolar2022} (heliocentric frame) & K/I & $0.01 - 1$ & $5 \pm 1.6 \times 10^{-3}$ & $252 \pm12$ & $65 \pm 9$ & $3.3\sigma$ \\
    TGSS radio galaxies~\cite{bengalyProbingCosmologicalPrinciple2018} & K/I & $0.01 - 4$ & $7.0 \pm 0.4 \times 10^{-2}$ & $243 \pm 12$ & $45 \pm 3$ & $99.5\%$ C.L.  \\
    NVSS radio galaxies~\cite{bengalyProbingCosmologicalPrinciple2018} & K/I & $0.01 - 4$ & $2.3 \pm 0.4 \times 10^{-2}$ & $253 \pm 11$ & $27 \pm 3$ & $99.5\%$ C.L.  \\
    quasars \cite{secrestTestCosmologicalPrinciple2021} & K/I & $0 - 3.6$ & $1.5 \times 10^{-2}$ & $238$ & $28$ & $4.9\sigma$ \\
    $\alpha$ \cite{wilczynskaFourDirectMeasurements2020} & I & $0.2 - 7.1$ & $0.72 \pm0.16 \times 10^{-5}$ & $325 \pm 17.5$ & $-11 \pm 10.3$ & $3.9\sigma$ \\
    $H_0$ via galaxy clusters \cite{migkasCosmologicalImplicationsAnisotropy2021} & I & $0.004 - 0.839$ & $9\%$ variation & $273 \pm 40$ & $-11 \pm 27$ & $5.4\sigma$ \\
    \hline
    \end{tabular}
    \label{tab:dipoles}
\end{table*}

We shall refer to anisotropies that cannot be entirely attributed to relative motion as ``intrinsic'' in this paper. Previous studies of intrinsic anisotropies include a dipole in the acceleration of the Universe~\cite{colinEvidenceAnisotropyCosmic2019}, a dipole in galaxy cluster scaling relations~\cite{migkasAnisotropyGalaxyCluster2018, migkasProbingCosmicIsotropy2020, migkasCosmologicalImplicationsAnisotropy2021}, and spatial variation in the fine-structure constant $\alpha$~\cite{kingSpatialVariationFinestructure2012,wilczynskaFourDirectMeasurements2020}. The aforementioned discrepancies in kinematic dipoles may likewise suggest an intrinsic dipole. While these intrinsic anisotropies may indicate a major deviation from $\Lambda$CDM at large scales, they could instead just be relatively-local anomalies since a given anisotropy can only be confirmed up to the maximum redshift of the astrophysical probe used. Regardless, local or low-redshift anisotropy can still contradict the CP and $\Lambda$CDM, thus warranting further investigations of both kinematic and intrinsic dipoles.

Some non-$\Lambda$CDM models can accommodate intrinsic anisotropies. For example, some Bianchi spacetimes allow for both homogeneity and anisotropy. These spacetimes have been well-studied, as they can reduce to a universe following the CP in the presence of a positive cosmological constant~\cite{ellisClassHomogeneousCosmologieal,waldAsymptoticBehaviorHomogeneous1983}. However, most of the Bianchi spacetimes have been shown to be incompatible with CMB observations, leaving only the Bianchi $\mathrm{VII}_h$ spacetime as a potential candidate \cite{aluriObservableUniverseConsistent2023}. But even the $\mathrm{VII}_h$ spacetime faces issues: while it is favored when attempting to fit only the CMB temperature map separately from $\Lambda$CDM, the best-fit Bianchi model possesses cosmological energy densities for matter ($\Omega_m$) and dark energy ($\Omega_{\Lambda}$) that strongly disagree with those in $\Lambda$CDM and cannot reproduce the CMB polarization map~\cite{planckcollaborationPlanck2015Results2016a,saadehHowIsotropicUniverse2016a, aluriObservableUniverseConsistent2023}. Thus, the Bianchi $\mathrm{VII}_h$ spacetime cannot entirely account for the same observations as the $\Lambda$CDM model, even if it is favored by CMB temperature anisotropies alone.

A more-recently proposed approach is a modification of the FLRW metric via the addition of an arbitrary 1-form~\cite{changCosmologicalModelLocal2013}, motivated by the considerations of a privileged axis within ``Very Special Relativity''~\cite{cohenVerySpecialRelativity2006}. The spacetime metric of this model is of the Randers type~\cite{randersAsymmetricalMetricFourSpace1941}, which is part of the Finsler spacetime, a generalized geometry that reduces to Riemannian geometry as a special case~\cite{bao2012introductionFinsler}. The addition of the 1-form generates a preferred axis, which would introduce a direction-dependent modification of the distance-redshift relation. This model has been explored using the luminosity distances of supernovae with four distance calibration methods~\cite{changConstrainingAnisotropyUniverse2014}, where the authors found no strong evidence of anisotropy. A similar approach was performed on a combined dataset of gamma ray burst (GRB) sources and supernovae~\cite{changConstrainingAnisotropyUniverse2014a}, which indicated a small but nonzero anisotropy along a preferred axis. Other work has considered similar extensions of Randers-Finsler cosmology to explore anisotropy in the fine-structure constant and the supernova distance-redshift relation~\cite{liUnifiedDescriptionDipoles2015,liSpatialTemporalVariations2017}; these studies found small, overlapping dipoles for each probe, but the dipole amplitudes did not agree as would be expected in the Randers-Finsler case. Another work performed cosmological constraints of two simple implementations of the Randers-Finsler spacetime using a combination of supernovae, baryon acoustic oscillations, and the temperature+lensing of the CMB~\cite{wangFirstComprehensiveConstraints2018}, finding that both models are consistent with the flat $\Lambda$CDM model.

To our knowledge, the Bianchi and Rander-Finsler cosmologies are the primary modern approaches for producing an anisotropic universe, and there are no other major candidates that can replace $\Lambda$CDM while accounting for the observed anisotropies. Given the null, minor, and/or inconclusive results for the models, we believe the current observations of intrinsic dipoles to still be unexplained by an underlying theory. However, a common prediction of several of the aforementioned models is a dipole in the expansion or distances in the Universe along an axis; we thus select a phenomenological model of anisotropy as a dipole in the luminosity distance.

The use of a luminosity distance dipole also circumvents a possible degeneracy between kinematic and intrinsic dipoles: while the two types of anisotropies have clearly distinct physical causes, they can produce similar observable effects. For example, when measuring luminosity distances across the sky in the presence of a dipole, a kinematic dipole would be degenerate with an intrinsic dipole in $H_0$ at low redshifts ($z \lesssim 1$) (see \cite{migkasCosmologicalImplicationsAnisotropy2021} for an explanation of this phenomenon using galaxy cluster luminosities). This is because an anisotropic value of $H_0$ would indicate anisotropic expansion of the Universe, which would alter measured luminosity distances through the redshift-distance relation (e.g., explored in the context of anisotropic dark energy in \cite{caiConstrainingAnisotropicExpansion2013}). On the other hand, a kinematic anisotropy or bulk flow would change the observed flux from an event, thus similarly altering the measured luminosity distances~\cite{bonvinFluctuationsLuminosityDistance2006,bonvinDipoleLuminosityDistance2006}. These two types of dipoles would hence have identical effects, and thus a dipole in luminosity distance at low redshifts could be attributed to either type of anisotropy~\cite{aluriObservableUniverseConsistent2023}.

An intrinsic dipole in luminosity distance has been suggested by studies of galaxy clusters, wherein a variation in the distance to a cluster can be measured by various scaling relations~\cite{migkasAnisotropyGalaxyCluster2018, migkasProbingCosmicIsotropy2020, migkasCosmologicalImplicationsAnisotropy2021}. These studies find a dipole-like anisotropy with a magnitude that deviates from the predictions of $\Lambda$CDM; due to the degeneracy of kinematic and intrinsic dipoles, it is not clear whether this dipole is due to a spatial variation in $H_0$ or a large-scale bulk flow. Other work has utilized standard candle supernovae to probe anisotropies, using the luminosity distance~\cite{caiConstrainingAnisotropicExpansion2013,andradeIsotropyLowRedshift2018} or distance modulus~\cite{bengalyjr.ProbingCosmologicalIsotropy2015,linSignificanceAnisotropicSignals2016,changTomographicTestCosmological2018,huTestingCosmicAnisotropy2020,tangTangConsistencyPantheonSupernovae2023}. The results of these studies tend to be slightly anistropic or consistent with isotropy, but they stress the caveat that the datasets used are greatly limited in redshift depth and sky completeness.

The current limitations of some electromagnetic studies of anisotropy can potentially be overcome by considering gravitational waves (GWs) from compact object binary mergers as another, independent probe. GWs from merger events can be used to probe both kinematic and intrinsic anisotropies because---just like astrophysical sources of electromagnetic waves---the amplitude of the observed signal depends on the luminosity distance and sky location of the source. In \cite{stiskalekAreStellarmassBinary2020}, the authors search for an intrinsic population anisotropy in the 2D spatial distribution of 10 GW events from the LIGO-Virgo O1 and O2 GW catalogs, finding that neither isotropy nor anisotropy is preferred. The authors of \cite{essickIsotropyMeasurementGravitational2023} use the LIGO-Virgo-KAGRA (LVK) O1-O3 catalogs to constrain the anisotropy in merger rates across the sky, setting upper limits on anistropy of $3.5\%-16\%$ at $90\%$ confidence, depending on the correlation length scale. Regarding a kinematic dipole coincident with the CMB dipole, \cite{kashyapDipoleAnisotropyGravitational2023} showed how the number count and chirp mass distributions of binary black hole (BBH) events will be impacted by Earth's motion; after applying this idea to BBHs found during the O1-O3 observing runs, they found insignificant support for anisotropy (p-value of $0.3$) and note that $\mathcal{O}(10^5)$ events would be required for a $3 \sigma$ detection of anisotropy.

All of these studies have found results that are generally consistent with isotropy, but they also note the obvious caveat that the current catalogs are extremely limited by the small number of current GW observations (<$100$) and the large position errors resulting from poor sky localization. Both of these factors are crucial for detecting a dipole of either type: the sky must be well-mapped by a large number of well-localized events to accurately measure the direction and magnitude of a dipole anisotropy.

One solution to poorly-localized GW events is to use ``bright standard sirens,'' events that have an observable electromagnetic counterpart. One example of a bright standard siren is a binary neutron star (BNS) merger that possesses an observable kilonovae or GRB. Since the distance to a GW event can be obtained using only the GW signal itself, bright standard sirens have various applications in cosmology (such as measuring the Hubble parameter) since the observation of a counterpart allows the event's redshift to be determined~\cite{schutzDeterminingHubbleConstant1986, holzUsingGravitationalWave2005,chenTwoCentHubble2018}. Additionally, the counterpart allows the event to be well-localized in the sky; if enough events are detected, the sky could be comprehensively and accurately mapped-out, which would bolster studies of anisotropy. Unfortunately, however, the current generation (CG) of GW detectors has so far discovered only one BNS event with a confirmed electromagnetic counterpart~\cite{abbottGW170817ObservationGravitational2017} and only several BNS kilonovae are expected to be observed with current/upgraded LVK detectors~\cite{shah2023predictions,colomboMultimessengerObservationsBinary2022,frostigInfraredSearchKilonovae2022,kiendrebeogoUpdatedObservingScenarios2023}. Thus, in order to use bright standard siren BNSs to study anisotropies, next-generation (XG) detectors may be required.

Two of the ground-based XG detectors that are being planned are the Cosmic Explorer (CE)~\cite{reitze2019cosmic} and the Einstein Telescope (ET)~\cite{punturoEinsteinTelescopeThirdgeneration2010}. CE would be similar in design to existing inteferometers, but instead would have $20$ or $40$-km arms, while ET would be an underground setup with three detectors oriented in an equilateral triangle, each with 10-km arms. Both detectors are projected to observe a tremendous number of BNS events: CE should observe a total of $\sim 300,000$ events annually and cover $\sim 80\%$ of events up to redshift $z=1$~\cite{evansHorizonStudyCosmic2021}, while ET is expected to observe $\mathcal{O}(10^5)$ events annually up to redshifts $z \sim 2-3$ \cite{maggioreScienceCaseEinstein2020}. When combined with existing or upgraded current-generation detectors, one or more XG detectors would offer greatly-improved event localization, as well as many other scientific benefits \cite{maggioreScienceCaseEinstein2020,evansHorizonStudyCosmic2021,perkinsProbingFundamentalPhysics2021,guptaNSBHnextgeneration2023,corsiMMAnextgeneration2024}.

XG detectors are also expected to greatly improve our ability to measure cosmic dipoles. Several studies have forecast their capabilities to measure a kinematic dipole coincident with the CMB dipole. In \cite{mastrogiovanniDetectionEstimationCosmic2023}, the authors used the number counts of BBH merger injections for a network of one Einstein Telescope and two Cosmic Explorers (\ECS), ignoring BNS mergers to avoid threshold effects from events that do not pass the GW signal-to-noise (SNR) detection threshold. The authors found that a so-called ``AGN'' dipole with an amplitude $\sim 5$-times greater than the CMB dipole (i.e., $6 \times 10^{-3}$) would be detectable at $>3\sigma$ with $10^6$ events, while a dipole with amplitude equal to the CMB dipole would not be detected with less than $\mathcal{O}(10^7)$ events. This study was extended by \cite{grimmCombiningChirpMass2023}, wherein the authors developed a framework to use the luminosity distance, chirp mass, and number counts to measure a kinematic dipole. After accounting for the threshold effects involved in detecting BNS mergers, the authors found that using both BBHs and BNSs from an \ECS network would allow the AGN dipole amplitude to be measured at $3\sigma$ with just $10^5$ events, while an amplitude of the CMB dipole could be measured at $\sim 1\sigma$ with $10^6$ events.

Additionally, other studies have made projections for measuring a dipole in the luminosity distance with single XG detectors. In \cite{caiProbingCosmicAnisotropy2018}, the authors considered bright standard sirens as observed by individual detectors: BNS and NSBH mergers with ET or the DECi-hertz Interferometer Gravitational wave Observatory~\cite{kawamuraJapaneseSpaceGravitational2006}, and supermassive BBH mergers as observed by LISA; the authors' results showed that ET observations of $\gtrsim 200$ events would favor anisotropy for a large dipole (amplitude $6 \times 10^{-2}$). Note that this result need not be consistent with the results of~\cite{mastrogiovanniDetectionEstimationCosmic2023} discussed above since the dipole considered in~\cite{caiProbingCosmicAnisotropy2018} is for the luminosity distance, not number count. In \cite{caiProbingCosmicAnisotropy2019}, the authors extended the previous work and considered BNS bright standard sirens that have joint fast radio burst observations for a single ET detector in order to measure a larger dipole (amplitude $0.1$), finding that this kind of standard siren would offer improved measurements of the dipole's amplitude but only comparable measurements of its direction.

To our knowledge, none of the current studies have considered several key ideas that are important for realistic dipole measurement forecasts with XG GW events. First, while studies use the sensitivities and observational parameters predicted for XG detectors, none use waveform models for the GW signal injections. Second, the detectors themselves are not simulated as being located at particular locations on Earth, and thus, the anisotropic sky sensitivity of each individual detector and detector network is omitted, as is the phase difference in the GW signal as it arrives at different detectors. Third, the motion of the detectors due to Earth's rotation is not considered, meaning that the sky sensitivity of the detectors is static. Fourth, the injected dipole is always considered to be at a fixed location; while this fiducial location is frequently chosen to be coincident with the CMB dipole, this choice ignores any location-dependence of the dipole constraint.

The last three points are especially important given that they involve the direction-dependence of detector networks, which is crucial for accurately measuring a dipole. For example, \cite{mastrogiovanniDetectionEstimationCosmic2023} does not account for sky direction dependence, but the authors note that it would result in a $\sim 3\%$ variation in SNR across the sky for an \ECS network, corresponding to a $\mathcal{O}(10^{-3})$ distortion of their projected measurements of the dipole amplitude. Given that most matter-based dipoles in recent studies have an amplitude of $\mathcal{O}(10^{-2})$ to $\mathcal{O}(10^{-3})$ (see Table \ref{tab:dipoles}), this consideration is likely to impact the accuracy of the projected constraints.

This paper addresses the four ideas described above using the GW detector simulation suite \texttt{gwbench}~\cite{borhanianGwbenchNovelFisher2021}. Our use of this software allows us to simulate BNS signal injections for networks of multiple ground-based CG and XG detectors with state-of-the-art GW models, realistic detector locations, and the time-evolution of detector sensitivity due to Earth's rotation. We consider a general dipole in the luminosity distance as a phenomenological model to allow for either a kinematic or intrinsic aniostropy. We also remain agnostic of whether the dipole affects GWs alone or both GWs and electromagnetic waves, since we use the electromagnetic counterparts of bright standard siren BNSs to obtain only the 2D angular sky localizations of the BNSs, not their distances or redshifts. We then simulate measurements of this dipole using BNSs as observed by six different detector networks, ranging from upgraded present-day detectors to the ideal case of multiple XG detectors (\ECS), while accounting for the fact that most BNS events will not have an observable electromagnetic counterpart. To assess the location-dependence of dipole constraints, we perform a simulation campaign with the \ECS network, wherein the dipole location is varied across a grid on the sky.

We find that for a luminosity distance dipole of amplitude $0.01$ and a year of observations, a network with a single XG detector would at best be able to measure the dipole at a level comparable to existing electromagnetic methods. With two XG detectors, these measurements become at best competitive or slightly superior to existing measurements; it is not until there are three XG detectors (\ECS) that the measurements are clearly superior. We also find that the directional sensitivity of a \ECS network does not play a particularly important role in the resulting dipole measurements. 

We present the details that support these findings in the remainder of this paper. We first discuss the modification and utilization of \texttt{gwbench} to simulate BNS signals and detector networks in Sec.~\ref{sec:method}. We describe our analysis of these simulations, present the resulting constraints on a luminosity distance dipole, and compare our results with existing dipole constraints in Sec.~\ref{sec:results}. We summarize our work and provide an outlook in Sec.~\ref{sec:conclusion}. Throughout the paper, we adopt the Planck 2018 cosmological parameters~\cite{planckcollaborationPlanck2018Results2020a} as implemented in \texttt{astropy.Planck18}~\cite{astropy2022} ($H_0 = 67.66 \,\mathrm{km / s / Mpc }, \, \Omega_{m,0} \approx 0.31,\, \Omega_{\Lambda,0} \approx 0.69$), utilize galactic coordinates in all Mollweide skymap plots, and assume that all CE detectors are configured with 40km arms and optimized for detecting compact binary mergers.
\section{Simulating gravitational-wave signals and detector networks}\label{sec:method}
We now describe how we create the mock signals of GW events (henceforth referred to as ``injections'') and estimate the accuracy to which the parameters of each injection can be estimated. Specifically, in Sec.~\ref{subsec:gwbench}, we describe the injection creation and parameter estimation code \texttt{gwbench} along with our modifications that support a dipole anisotropy. In Sec.~\ref{subsec:params}, we detail our process of modeling individual injections of BNS bright standard sirens and our procedure for generating populations of injections that serve as possibly-observable GW events. In Sec.~\ref{subsec:6netsim}, we introduce the emulated GW detector networks and discuss their mock observations of the injections.

\subsection{\texttt{gwbench} usage}\label{subsec:gwbench}
The \texttt{gwbench} software~\cite{borhanianGwbenchNovelFisher2021} allows for the study of GW event injections for a range of source masses, redshifts, and sky positions, for a variety of CG and XG detector networks. The emulated detectors themselves are configurable and allowed to be at a range of physical locations and orientations across Earth, providing a scientifically-realistic rendition of the XG era. \texttt{gwbench} includes Fisher information analysis to estimate the accuracy at which event parameters can be inferred from a given GW observation, wherein each parameter's posterior is assumed to have a single multidimensional Gaussian distribution (which is a reasonable approximation in the high SNR limit~\cite{Vallisneri:2007ev}). By estimating the posterior covariance matrix through a Fisher analysis, \texttt{gwbench} is a means to approximate measurement errors on model parameters---such as the luminosity distance---of future GW events, as well as the SNR of the event for each detector.

We modified \texttt{gwbench} to introduce a dipole anisotropy in the luminosity distance of the injections. The full modifications are contained in a Gitlab fork of the original code~\cite{gwbench-bfc}; the primary addition is the computation of a modified luminosity distance $D_L'$ based on a configurable dipole model:
\begin{align}
    D_L'(z,\hat{z}) = D_L^0(z) [1 + (\hat{n} \cdot \hat{z}) g]\label{eq-dipole}.
\end{align}
Here, $D_L^0(z)$ is the unmodified, isotropic luminosity distance of the injection, which is a function of only the redshift, $\hat{z}$ is the redshift unit-vector direction toward the injection, $g$ is the amplitude of the dipole, and $\hat{n}$ is the direction to the dipole, defined as:
\begin{align}
    \hat{n} = \left( \cos\phi\sin\theta, \sin\phi\sin\theta, \cos\theta \right)\label{eq-n}.
\end{align}
The angular coordinates ($\phi,\theta$) are simply spherical polar coordinates, which correspond to astronomical coordinates ($\phi$ is Right Ascension; $\theta$ is Declination+$90\degree$). Note that $D_L'$ is not a function of $(\phi, \theta, g)$ since the dipole parameters are held constant across different injections, leaving $\hat{z}$ as the only variable between injections. While we use a range of dipole directions between separate simulations, we keep the dipole amplitude fixed at $g=0.01$, based on the amplitudes of currently-measured matter-based dipoles discussed in Sec.~\ref{sec:intro} and shown in Table~\ref{tab:dipoles}. 

\subsection{Modeling binary neutron star mergers}\label{subsec:params}
We employ a general methodology in this paper to create single synthetic injections, to study the accuracy to which injection parameters can be measured, and to construct many injections to serve as a population of events that may be detected in the future with different networks of detectors. All of our studies require that the sky localization of each GW event be known; we thus exclusively use BNS mergers as bright standard sirens due to their potential to be accurately localized using observable counterparts, such as kilonovae. As such, we do not consider binary black hole mergers or other merger scenarios. Moreover, we assume an electromagnetic counterpart only to ensure the 2D sky localization of the merger---not to measure the merger's redshift---and thus do not consider redshift-determination methods such host galaxy identification.

We define a \textit{signal injection} as an individual, synthetic BNS event, generated by evaluating an inspiral waveform model at a given set of source parameters. Each event is modeled through the IMRPhenomPv2NRTidalv2 waveform model~\cite{dietrichMatterImprintsWaveform2019, dietrichImprovingNRTidalModel2019a}, which allows us to accurately capture the full inspiral, up to binary contact, while allowing for precession and tidal effects. 

We define the \textit{simulation set} for a specific detector network to be a population of BNS signal injections---with model parameters drawn from particular distributions---that constitutes the subset of events that are actually observable by a given detector network.
We construct a simulation set as follows. First, we draw samples from a set of chosen distributions for the source parameters of the injections (as summarized in Table~\ref{tab:parameters}). In particular, we sample component masses from a Gaussian function, spanning $[1, 2.5] M_{\odot}$ (with mean $\mu = 1.35 M_{\odot}$ and standard deviation $\sigma = 0.15 M_{\odot}$). We assume zero spin, $\chi_1 = 0 = \chi_2$, and set the time and phase of coalescence to zero, $t_c = \varphi_c = 0$. We sample the position parameters $\alpha$ (right ascension) and $\delta$ (declination) uniformly in $\alpha \subset [0, 2\pi]$ and $\cos(\delta) \subset [-1,1]$. We also uniformly sample the orbital inclination $\iota$ in $\cos(\iota) \subset [-1,1]$ and the GW polarization in $\psi \subset [0, 2\pi]$. Regarding tidal effects, the BNS waveform is only weakly affected by tides before merger. Therefore, it is only weakly correlated with the distance and does not affect the Fisher estimates of the luminosity distance measurements, which is our primary quantity of interest. So, for simplicity, we fix the individual component parameters $\Lambda_1$ and $\Lambda_2$ by assuming a constant value of the reduced tidal deformability $\tilde{\Lambda}$. $\tilde{\Lambda}$ is a linear combination of $\Lambda_1$ and $\Lambda_2$ which represents the total deformability of the merger~\cite{wade2014systematic}; we select a value of $\tilde{\Lambda}=100$ to be consistent with existing tidal constraints~\cite{abbott2019properties}.

\begin{table}[htp]
    \caption{Sampling parameters for the BNS injections and their respective priors. See the main text for an explanation of each parameter.}
    \begin{tabular}{|c|c|}
        \hline
        Parameter & \multicolumn{1}{c|}{Prior Distribution} \\
        \hline \hline
        $m_1$, $m_2$  & Gaussian  ($\mu=1.35 \, \pm 0.15\,M_\odot$) \\ \hline
        $\chi_1, \chi_2$ & 0 \\ \hline
        $z$  & Madau-Dickinson up to $0.5$ \\ \hline
        $D_L$  & convert $z$ using \texttt{astropy.Planck18} \\ \hline
        $\alpha$ & uniform in $[0,2\pi]$ \\ \hline
        $\cos(\delta + \pi/2)$ & uniform in $[-1,1]$ \\ \hline
        $\cos(\iota)$  & uniform in $[-1,1]$ \\ \hline
        $\psi$ & uniform in $[0,2\pi]$ \\ \hline
        $t_c$, $\varphi_c$ & $0$ \\ \hline
        merger time & uniform in [0, 86400] seconds \\
        \hline
    \end{tabular}\label{tab:parameters}
\end{table}

\begin{figure}[htp]
  \includegraphics[width=\columnwidth]{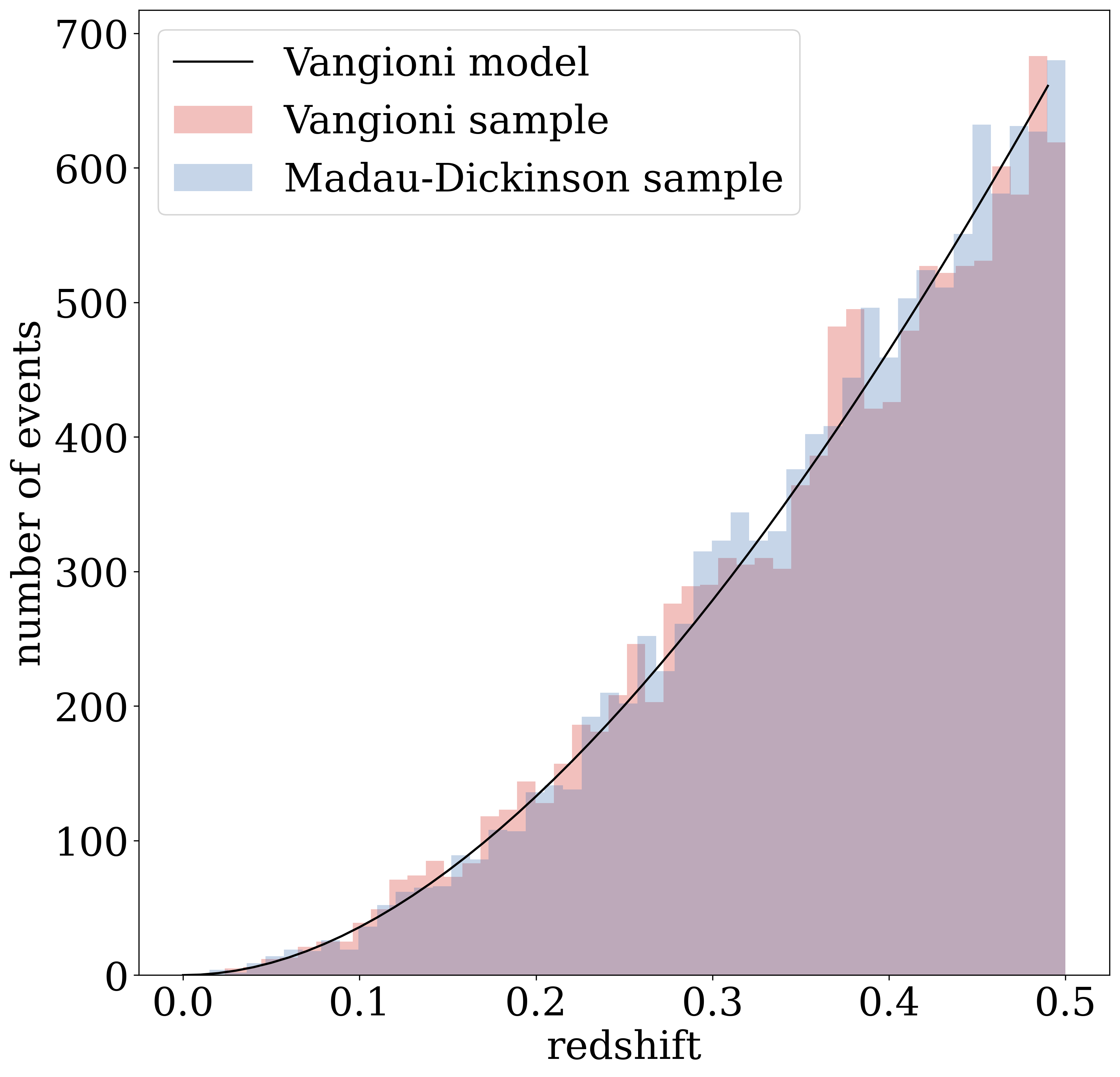}
  \caption{Comparison of the Madau-Dickinson (blue) and Vangioni (red) population models as a function of redshift $z$. The Madau-Dickinson sample was generated using the sampler implemented in \texttt{gwbench}, while the Vangioni sample was generated from the model described in~\cite{vangioniImpactStarFormation2015}, as implemented in~\cite{gwbench-bfc}. For comparison, we also plot the Vangioni model (black) that produced the sample. Observe that the samples do not differ appreciably in this redshift range.}
  \label{fig:pop-models}
\end{figure}

We define the the distribution of redshifts for each of the BNS injections by using standard astrophysical population models. The Madau-Dickinson (MD) population distribution~\cite{madauCosmicStarFormationHistory2014} is a well-established description of star formation rates and redshift distributions. To account for the passage of time between binary formation and merger, it is common to use a time-adjusted MD model wherein the time (redshift) of the merger is offset from the original MD distribution by a time delay $t_d$ drawn from $p(t_d) \propto t_d^{-1}$. We thus use a time-delayed version of the MD model, implemented within \texttt{gwbench} as the BNS redshift probability distribution function with the following form:
\begin{equation}
p_{BNS}(z) = \phi_0 \frac{ (1+z)^{a_0} } {1 + [(1+z)/c_0]^{b_0} }
\end{equation}
with parameters $a_0=1.8032$, $b_0=5.3098$, $c_0=2.8372$, $\phi_0=8.7659$, as used in~\cite{borhanianListeningUniverseNext2022}. We use this parameterized MD model in a redshift range of $z \subset[0.001, 0.5]$, the upper bound being due to the anticipated limits of EM follow-up observations for kilonovae~\cite{chaseKilonovaDetectabilityWidefield2022,borhanianListeningUniverseNext2022}. To cross-validate this usage of a population model, we compared some of our results to those obtained with the more-recent Vangioni model~\cite{vangioniImpactStarFormation2015} during preliminary analyses and found that it produced a nearly-identical population as the MD model in our redshift range (shown in Fig. \ref{fig:pop-models}). Using a redshift drawn from this MD distribution, we then compute the event's isotropic luminosity distance $D_L^0$, using the Planck 2018 cosmological results~\cite{planckcollaborationPlanck2018Results2020a}; this $D_L^0$ is then modified in accordance with Eq.~\eqref{eq-dipole} to obtain the anisotropic $D_L'$, which is used to create each injection in each simulation set. 

Although we modify the luminosity distance for each of the injections we consider, we assume the default IMRPhenomPv2NRTidalv2 waveform model of~\cite{dietrichMatterImprintsWaveform2019, dietrichImprovingNRTidalModel2019a}---without dipole modifications---to recover the distance. In practice, this is implemented within \texttt{gwbench} as a Fisher estimate of the accuracy to which the IMRPhenomPv2NRTidalv2 model parameters can be measured. The model parameters that are estimated are then $\{ \eta, \log(D_L), t_c, \phi_c, \cos(\iota), \psi \}$. Notice, in particular, that $g$ and $\hat{n}$ are not model parameters, as these would be entirely degenerate with $D_L^0$. The Fisher matrix is hence $6\times6$ dimensional for each injection, and the square root of the diagonal of its inverse provides a marginalized estimate of the accuracy to which the luminosity distance---and other waveform parameters---can be measured.

When observing GW events, the time of year dictates the location of an event on the sky relative to the detectors that observe the event. CG ground-based detectors are sensitive to BNS signals only for tens of minutes, and thus, time evolution is not critical when considering such individual signals. However, BNS signals can last several hours or even days in the sensitivity band of XG detectors (see, e.g., Fig. 2 in~\cite{punturoEinsteinTelescopeThirdgeneration2010}), and for such signals the temporal evolution of the detector network is important. 
Moreover, detector temporal evolution must be taken into account when considering many distinct signals occurring over time across the sky. Thus, to accurately model a population of BNS signals, we consider a day-long time evolution of the detector networks to reflect the rotational motion of the Earth. We selected this window instead of a year-long window because the numerical differentiation of injection timestamps within \texttt{gwbench} is limited by floating point precision, so a timestamp range of a year may cause inaccurate results. Choosing a day-long window does not impact our results because it accounts entirely for Earth's \textit{rotation}, which is the primary effect of time-evolution; Earth's \textit{revolution} only introduces an error of order micro-arcseconds (for redshifts $0.001 < z < 0.5$) due to neglecting the parallax effect when localizing the GW event. This error is much smaller than the resolution of even XG networks; moreover, we later subsample the injections to account for a localization estimate of $1 \, \mathrm{deg}^2$ for electromagnetic follow-up (discussed below). Thus, a day-long window is functionally equivalent to a year-long window\footnote{Note: \texttt{gwbench} had a bug related to the computation of the merger time that was patched in \texttt{v0.8.0} after our simulations, as noted in \href{https://gitlab.com/sborhanian/gwbench/-/blob/master/CHANGELOG.md}{https://gitlab.com/sborhanian/gwbench/-/blob/master/CHANGELOG.md}. However, the bug would cause incorrect results only when Earth's rotation is not enabled and the merger time is zero; thus, our simulations were not impacted since we sample the merger time randomly and enabled Earth's rotation. We verified this empirically using two diagnostic simulations.}. Even with this choice, however, other differentiation errors still occur for $< 1\%$ of injections when the mass ratio $\eta$ is close to the derivative step size $\Delta$: $|\eta - 0.25| < \Delta$. For these injections alone, we change the numerical differentiation method from a \texttt{central} difference to a \texttt{backward} difference.

When creating simulation sets for different networks, we must ensure that the underlying time-evolving populations of injection are identical. To this end, we create each population of injections from the same ``seed.'' This effectively ensures that the $n^{\mathrm{th}}$ injection always possesses the same timestamp for each network, while allowing that injection to possess a different timestamp than the other injections. Thus, each network observes the same time-evolving injection population, leaving the detectors and dipole location as the only independent variables that can change between simulations.

Given this setup, we then create $12,000$ BNS signal injections to emulate a year-long dataset, in accordance with estimates of the yearly merger rates in our redshift range (Table VII in \cite{borhanianListeningUniverseNext2022}). Not all of these events, however, would be observable by our detector networks. We impose a network signal-to-noise ratio (SNR) threshold of $\rho \geq 10$ to determine which of the individual BNS injections can be considered as ``detected'', in agreement with the $\rho \sim 8-10$ threshold used in current LVK catalog papers~\cite{GWTC2CompactBinary2021,GWTC3CompactBinary2021}. Out of these detections, only a few would realistically receive an electromagnetic counterpart, which we need in our analysis to provide a measurement of their localization. Indeed, only $\sim 15\%$ of BNS events are expected to be localized to within $1 \, \mathrm{deg}^2$ prior to follow-up for an \ECS network (Table VII in~\cite{borhanianListeningUniverseNext2022}). Since our signal injections are assumed to have no localization error, we use SNR as a proxy for EM follow-up. Because events with higher SNRs are more likely to be well-localized~\cite{mageeRealisticObservingScenarios2022,borhanianListeningUniverseNext2022}, we select only the injections with the top $15\%$ highest SNRs and assume that $100\%$ of these events would have detected EM counterparts. All of the BNS injections that pass these cuts then form a simulation set for the given detector network.

Note that this $15\%$-highest SNR cut is just a proxy for EM followup, which---while physically-motivated---is not exact since in practice, even a well-localized high-SNR event might, e.g., lie beyond the survey volume of EM telescopes (making its kilonova undetectable) or be off-axis (so its gamma ray burst cannot be observed). Thus, this selection will result in an optimistic estimate. Moreover, the $\sim 15\%$ projection is given for an \ECS network, whereas less-sensitive networks may not achieve the fraction of well-localized events. However, as discussed later in Sec.~\ref{sec:results}, our results for these networks can simply be scaled by $1/\sqrt{N}$ for $N$ events that actually receive follow-up; this scaling could likewise be applied in the possible scenario where $15\%$ is not an accurate estimate for an \ECS network.

\begin{figure*}[!htp]
  \centering
  \includegraphics[width=\textwidth]{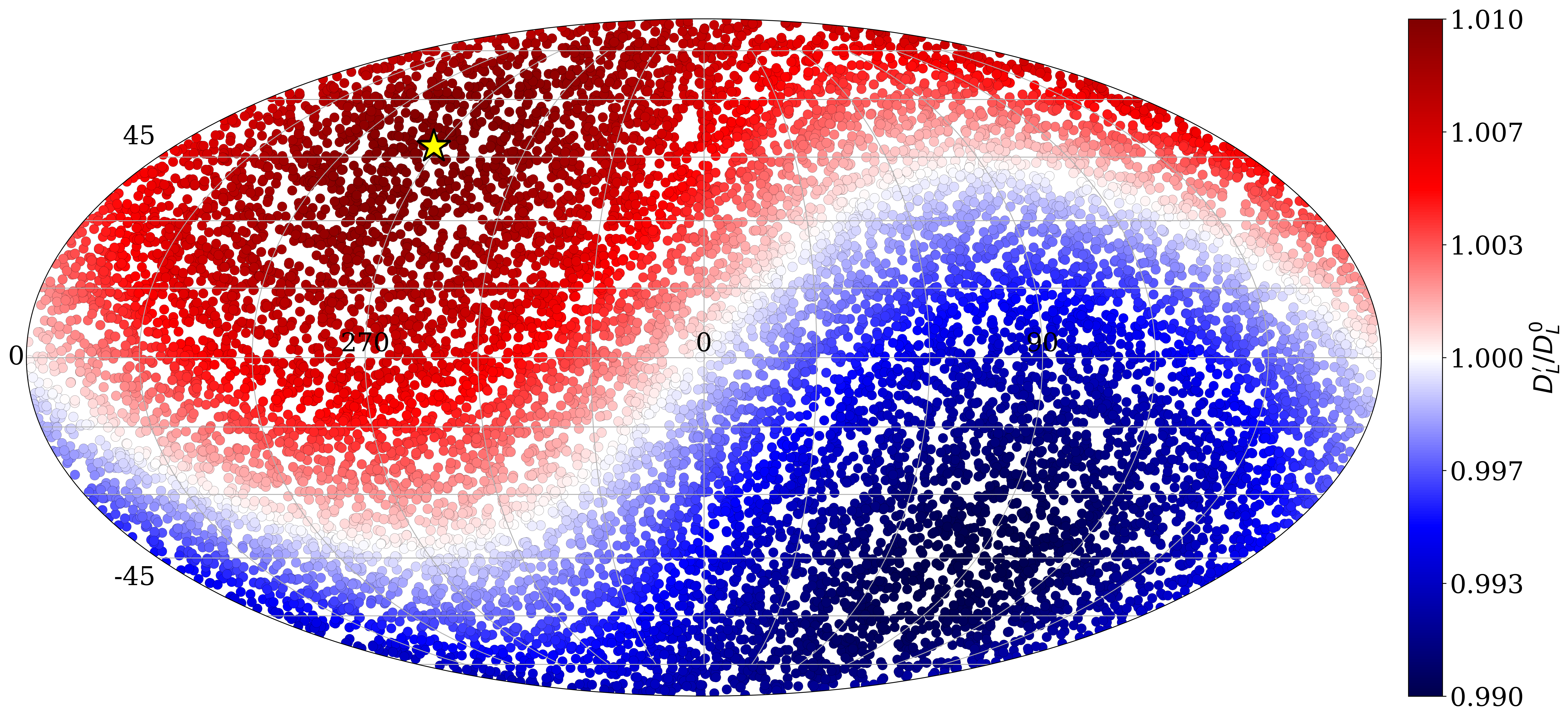}
  \caption{The anisotropic luminosity distances for a simulated dataset of 12,000 BNS injections. The yellow star indicates the location of the inserted dipole of amplitude $g=0.01$, which modifies the observed luminosity distance $D'_L$ relative to the original, isotropic luminosity distance $D_L^0$, as shown in the colorbar. Distances parallel to the dipole direction are increased (red) and distances antiparallel to the dipole direction are decreased (blue), while orthogonal distances are unchanged (white). Note that the distribution of injections is isotropic across the sky since we consider a generic dipole in distance, whereas a population anisotropy or kinematic dipole would induce an anisotropic distribution of injections.}
  \label{fig:dl-mw}
\end{figure*}

\subsection{Modeling GW detector networks}\label{subsec:6netsim}
We consider a variety of networks, each with a corresponding simulation set. The first simulation sets involve six networks consisting of advanced-plus LIGO, Voyager, and/or XG detectors, all with a fixed dipole location. The later simulation sets consist of a single network of three XG detectors, but with the dipole direction varying across the sky. Henceforth, we specify different detectors and networks following the convention of \texttt{gwbench} (Table 3 of~\cite{borhanianGwbenchNovelFisher2021}):
\begin{itemize}
\setlength\parskip{0.0cm}
\setlength\parsep{0.0cm}
    \item \textit{Current Generation}: LIGO Hanford (\texttt{H}), Livingston (\texttt{L}), and India (\texttt{I}); Virgo (\texttt{V}); KAGRA (\texttt{K})
    \item \textit{Upgraded Current Generation}: advanced-plus (\texttt{+}) and Voyager (\texttt{Voy})
    \item \textit{Cosmic Explorer}: northern location (\texttt{C}) and southern location (\texttt{S})
    \item \textit{Einstein Telescope}: (\texttt{E})
\end{itemize}

We then assemble these detectors into six networks, outlined in Table~\ref{tab:networks}. Regarding the CG detectors and their upgrades, both LIGO India~\cite{unnikrishnanIndIGOLIGOIndiaScope2013,saleemScienceCaseLIGOIndia2022} and KAGRA may be operational at design-sensitivities by the 2030s, and thus we include them with the \texttt{HLV}+ to form the \aplus network. Voyager~\cite{VoyagerAdhikariCryogenicSiliconInterferometer2020} is the name of a proposed cryogenic upgrade to LIGO Hanford/Livingston/India that would take place after 2030 and would thus replace those three detectors, with Virgo and KAGRA continuing to operate at advanced-plus sensitivity in the \voy network. We then consider four XG networks expected to be accessible after 2035 in order to compare the scientific potential of the different possible XG detectors. We first consider two networks, each with one of the two XG detectors: \et wherein Einstein Telescope replaces Virgo in Europe, and \cefo with a single Cosmic Explorer operating in place of the U.S. LIGO sites. We then consider combinations of XG detectors: \EC with the Einstein Telescope and the U.S. Cosmic Explorer, and \ECS with the Einstein Telescope and both Australian and U.S. Cosmic Explorers~\cite{evansHorizonStudyCosmic2021}. 

\begin{table*}[htp!]
    \caption{Labels and descriptions for the six networks of CG and XG detectors considered in this study, in order of increasing detection sensitivity. Labels follow the \texttt{gwbench} conventions detailed in Table 4 of~\cite{borhanianGwbenchNovelFisher2021}. Note: all Cosmic Explorer detectors are considered to be configured with 40 km arms and optimized for detecting compact objects.}
    \begin{tabular}{|c|c|}
        \hline
        Network Label & \multicolumn{1}{c|}{Detectors} \\
        \hline \hline
        \aplus & five CG\texttt{+} \\ \hline
        \voy & Voyager upgrades to three CG and two CG\texttt{+} \\ \hline
        \et & Einstein Telescope with four CG\texttt{+} \\ \hline
        \cefo & Cosmic Explorer with three CG\texttt{+} \\ \hline
        \EC & Cosmic Explorer, Einstein Telescope, and two CG\texttt{+} \\ \hline
        \ECS & Einstein Telescope and two Cosmic Explorers\\ \hline
    \end{tabular}\label{tab:networks}
\end{table*}

We now present the simulation sets for the six different networks described above. In these simulation sets, the dipole anisotropy has a fixed location for all networks; the location was set to be the same as the CMB solar dipole \cite{planckcollaborationPlanck2018Results2020}: ($l=268 \degree$, $b=48 \degree$) in galactic coordinates, ($\alpha=170\degree$, $\delta=-9\degree$) in ICRS coordinates, or ($\phi=170 \degree$; $\theta=81 \degree$) in the coordinates of Eq.~\eqref{eq-n}. The effect of this dipole on the luminosity distances of the full injection population for this simulation is shown in Fig. \ref{fig:dl-mw}. Observe that the luminosity distances are enlarged in the direction of the dipole and suppressed in the opposite direction, as expected. 

\begin{figure}[htp]
  \centering
  \includegraphics[width=1\columnwidth]{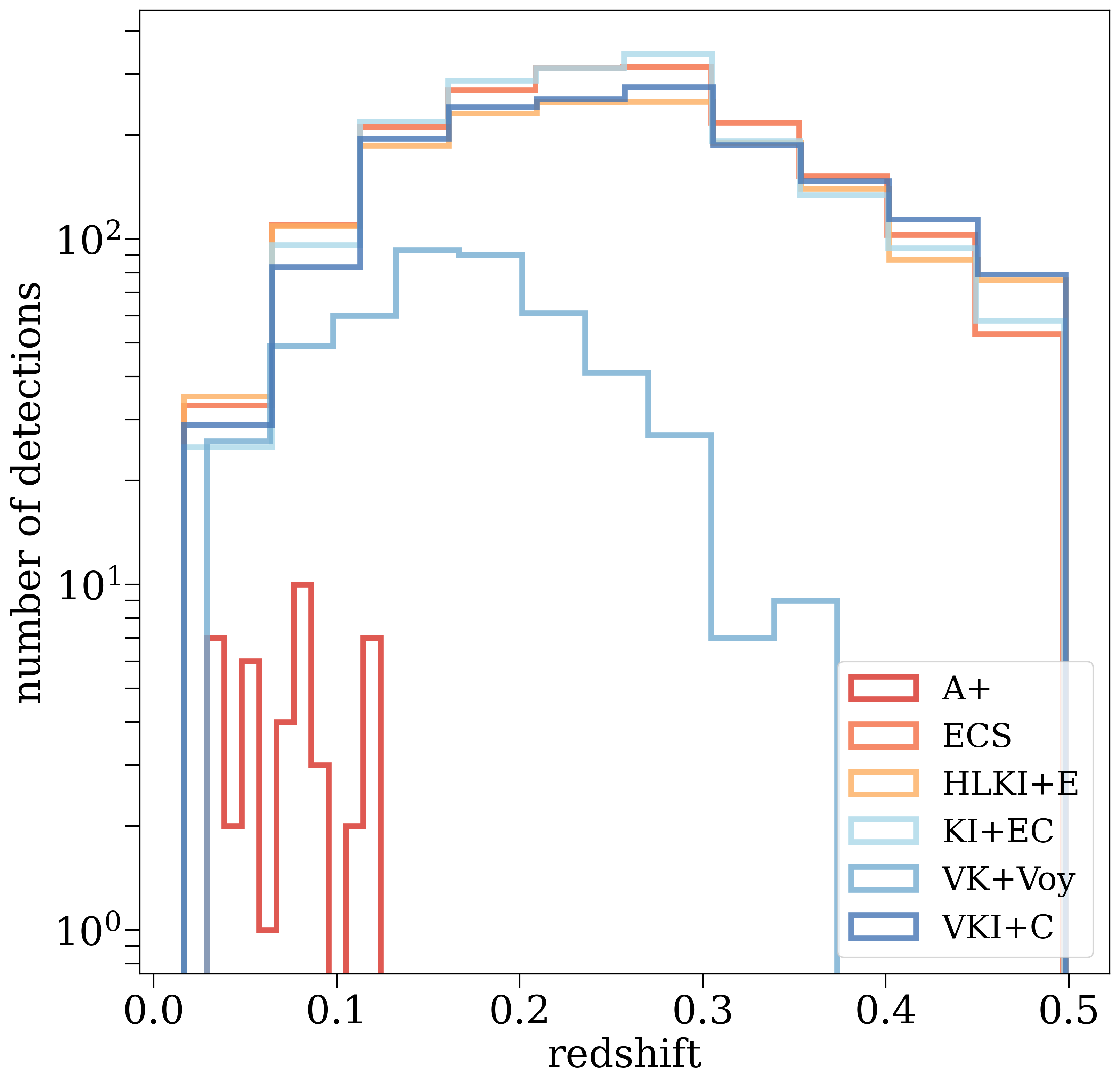}
  \caption{Redshift distributions of the injection population for each network after an SNR cutoff of 10. The colors indicate the network, as shown in the legend. Observe that the networks with at least one XG detector (red, orange, cyan, and dark blue) detect similar numbers of injections up to redshift $z = 0.5$, while the CG network (yellow) detects far fewer injections even at lower redshifts.}
  \label{fig:aniso_z}
\end{figure}

While each network is provided with the population of $12,000$ BNS signal injections, not all of these injections would be observed by all detectors at the same level of significance. Indeed, the $\rho \geq 10$ threshold results in different numbers of confirmed detections, as shown in Fig.~\ref{fig:aniso_z}. In more detail, this figure shows the redshift distributions of the injections within each network's simulation set. Observe that the more-sensitive XG networks are able to detect the majority of the $12,000$ injections up to a redshift of $z = 0.5$, the upgraded CG network is sensitive only to a fraction of injections up to $z \sim 0.37$, and the CG network is not sensitive to injections beyond $z \sim 0.12$. Note that the results for the \cefo and \et networks suggest that CE has slightly greater sensitivity than ET at all redshifts in this range. 

\begin{figure}[htp!]
  \includegraphics[width=1\columnwidth]{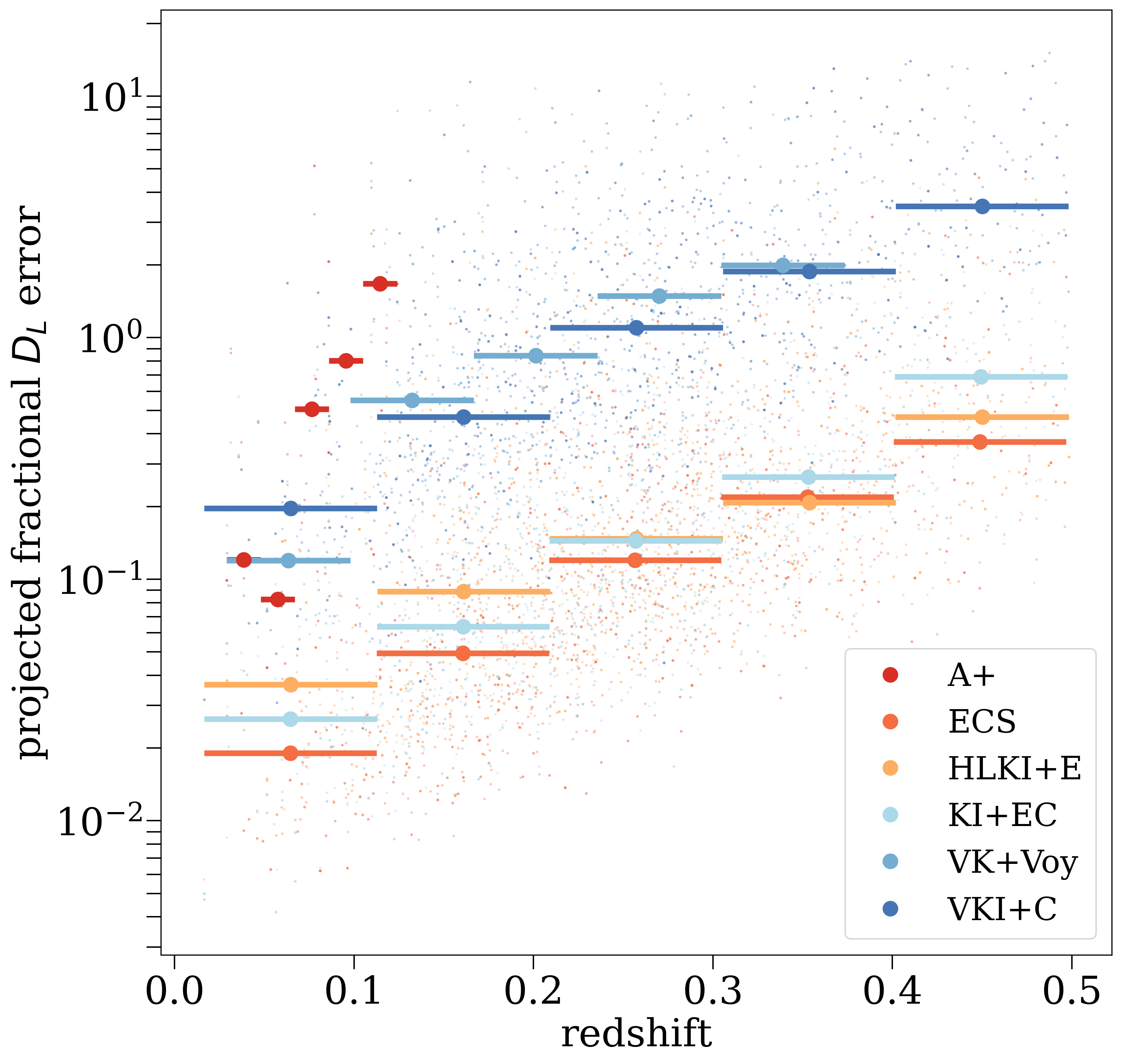}
  \caption{Errors on the luminosity distance measurements as a function of redshift for each network after an SNR cutoff of 10. For visualization purposes, the injections for each network are put into five bins whose medians and widths are shown by dark horizontal circles and lines, respectively. The less-sensitive networks (yellow and light blue) generally possess higher errors and fewer events at lower redshifts as compared to the more-sensitive networks (orange, cyan, red), but see the text for a discussion about the \cefo network (dark blue).}
  \label{fig:aniso_dl}
\end{figure}

The variety of distributions for these different simulation sets is mirrored in the distributions of luminosity distance measurement errors, as shown in Fig.~\ref{fig:aniso_dl}, as obtained via the Fisher information procedure. To demonstrate the average tendencies of each network, we separate the simulation sets into five redshift bins that are of equal width for a given network, and display the median of the $D_L$ fractional error for injections within that bin. Observe that the simulation sets of less-sensitive networks are smaller and possess higher errors as compared to the more-sensitive networks shown lower in the plot. Note also that, overall, the farther the system is, the more difficult it is to accurately measure the luminosity distance, as expected. Observe finally that our ability to measure luminosity distance is worse for the \cefo network containing one CE versus the \et network containing one ET, in contrast to the \cefo network's superior capability for injection detection noted in Fig.~\ref{fig:aniso_z}. This is because the ET observatory is planned to contain multiple co-located XG interferometers in its design; each interferometer can then measure a different GW polarization, since the interferometers are not co-aligned~\cite{punturoEinsteinTelescopeThirdgeneration2010,maggioreScienceCaseEinstein2020}. This allows for a more accurate measurement of the merger's inclination angle, which breaks the luminosity-distance/inclination angle degeneracy and results in a more precise measurement of the luminosity distance~\cite{dalalShortGRBBinary2006,usmanConstrainingInclinationsBinary2019}. 

While this strategy of addressing the distance--inclination degeneracy works for any detector network, it relies on the capability of at least two detectors, and thus it applies only to the redshift range that is covered by two or more detectors. Hence, a network with one CE would accurately measure inclinations---and luminosity distances---only up to the maximum redshift sensitivity of the second-most-sensitive detector in the network (in this case, LIGO+). This is much more limited than a network with one ET, which effectively contains multiple XG interferometers. Due to the need to rely on the relatively-limited redshift coverage of LIGO+, the \cefo network only manages to obtain luminosity distance errors of the same scale as the non-XG, Voyager-only network, albeit with many more detections up to a higher redshift range.

\begin{figure}[H]
  \centering
  \includegraphics[width=0.8\columnwidth]{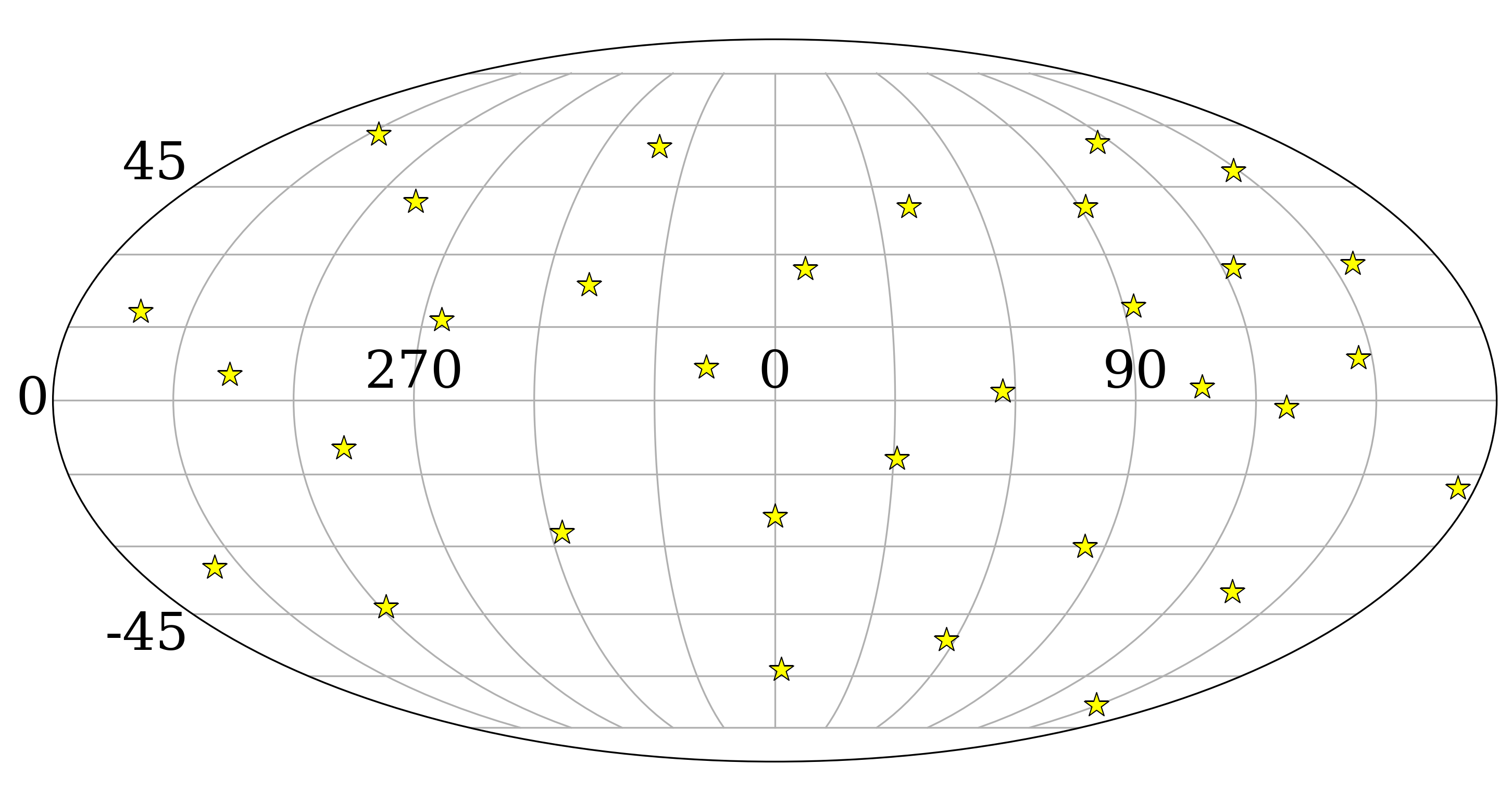}
  \caption{Grid of dipole locations used during the suite of \ECS network simulations. Note that two dipole locations were excluded from further analysis to avoid distortions arising from proximity to coordinate singularities near the galactic poles.}
  \label{fig:lb_grid}
\end{figure}

To assess the dependence of our analysis on the direction of the dipole, we also perform a suite of simulations wherein the dipole direction is varied across a 32-point grid for the \ECS network. We then interpolate over this grid to provide projected constraints for all possible locations of a dipole. 
To ensure valid interpolation, we select the grid spacing based on the \ECS network's $3\sigma$ errors on the dipole location in $(\theta, \phi)$ coordinates, obtained by analyzing the six-network simulation after the EM follow-up cutoff (see Sec.~\ref{subsec:6net} for a full description of these results). We determined the grid point spacing by:
\begin{itemize}
\setlength\parskip{0.0cm}
\setlength\parsep{0.0cm}
    \item treating the $1\sigma$ errors as radii of an ellipse: $\sigma_\theta \sim 7 \degree $ and $\sigma_\phi \sim 8 \degree$,
    \item scaling the radii to $3\sigma$ and computing the area of the error ellipse as $A_{error}\mathrm{~}\sim 1600 \,\, \mathrm{deg}^2$,
    \item computing the minimum number of grid points required to cover the sky: ${A_{sky}}/{A_{error}} \sim 26$.
\end{itemize}
To be conservative, we rounded the number of grid points up to $32$. We generated the grid to be uniform in $[\alpha,\,\cos(\delta)]$-space and then added two additional grid points, one at each of the poles ($\delta = 0, \pi$), to ensure valid interpolation. Then, due to issues with coordinate singularities, we omitted two grid points that were located near the galactic poles after transforming from equatorial coordinates. The resulting grid points are shown in Fig.~\ref{fig:lb_grid}. Note that any supposed distortions in the grid are entirely due to the transformation to galactic coordinates and the projection effects of the Mollweide display; the grid nonetheless has fair coverage of the sky.

\section{Projected constraints on a luminosity distance dipole}\label{sec:results}
Here, we analyze the simulation sets described in the previous section to constrain a dipole in the luminosity distance. In subsection~\ref{subsec:chi2}, we describe the three-parameter $\chi^2$ minimization procedure we use to constrain the dipole anisotropy. In the remaining subsections, we provide the results for each simulated detector network. 

\subsection{$\chi^2$ statistic}\label{subsec:chi2}

Similar to \cite{caiProbingCosmicAnisotropy2018}, we construct a $\chi^2$ statistic to determine the best-fit parameters for the dipole model described in \eqref{eq-dipole}, for each simulation set:
$$
\chi^2 = \sum_{i=1}^N \left[ \frac{D_L^i - D'_L(g_f,\hat{n}_f,\hat{z}_i)}{\sigma_{D_L^i}} \right]^2.
$$ 
Here, $N$ is the number of (detected) injections in the simulation set, while $D_L^i$ and $\hat{z}^i$ are the $i^\mathrm{th}$ injection's luminosity distance and redshift direction, while $\sigma_{D_L^i}$ is the $1\sigma$ error in the measurement of $D_L^i$ as estimated from a Fisher analysis carried out with \texttt{gwbench}. The function $D'_L$ is the anisotropic luminosity distance prescribed through the dipole model of Eq. \eqref{eq-dipole}, which is a function of the fitting parameters $g_f$ and $\hat{n}_f$. The $\chi^2$ statistic can then be minimized by varying ($g_f$, $\hat{n}_f(\theta,\phi)$) to find the best fit values of these parameters given the distribution of luminosity distances and their estimated uncertainties in a simulation set. We here employ the Nelder-Mead algorithm~\cite{nelder1965simplex} as implemented\footnote{\href{https://docs.scipy.org/doc/scipy/reference/optimize.minimize-neldermead.html}{https://docs.scipy.org/doc/scipy/reference/optimize.minimize-neldermead.html}} in SciPy~\cite{virtanen2020scipy} to minimize this $\chi^2$ statistic in the 3-dimensional ($g_f$, $\hat{n}_f(\theta,\phi)$) space.

To construct distributions of possible dipole measurements, we repeat the minimization process for $M=1000$ realizations of the simulation set. We utilize this approach in order to emulate $M$ possible ways a given detector network could observe a particular simulation set. We construct the realizations as follows. For each injection, we create a set of $M$ values of $D_L^i$. Each value is obtained by sampling from a Gaussian distribution with a mean of $D'_L(g_0, \hat{n}_0, \hat{z}_i)$ computed using Eq.~\eqref{eq-dipole} with true dipole parameters $(g_0, \hat{n}_0$) and a standard deviation of $\sigma_{D_L^i}$, for each injection. We apply the $\chi^2$ procedure to each of the realizations to obtain $M$ sets of the best-fit dipole parameters $(g,\phi,\theta)$. We then report the mean and $1\,\sigma$ standard deviation of these sets in galactic coordinates as ``constraint forecasts.''

\subsection{Six-network simulation}\label{subsec:6net}

After performing the SNR cutoff and EM follow-up subselection on the dataset of injections (Fig. \ref{fig:dl-mw}) to account for detecting bright standard sirens, we apply the $\chi^2$ procedure to the remaining injections. The resulting constraint forecasts, along with the number of bright standard sirens used in each analysis, are summarized in Table~\ref{tab:6net-all}. Regarding the number of detections, note that the CG\texttt{+} network detects less than 50 events; Voyager upgrades are required to get hundreds of detections. Detecting $\mathcal{O}(10^3)$ events is not possible until at least one XG detector is added.
\begin{figure*}[ht]
  \centering
  \includegraphics[width=1\textwidth]{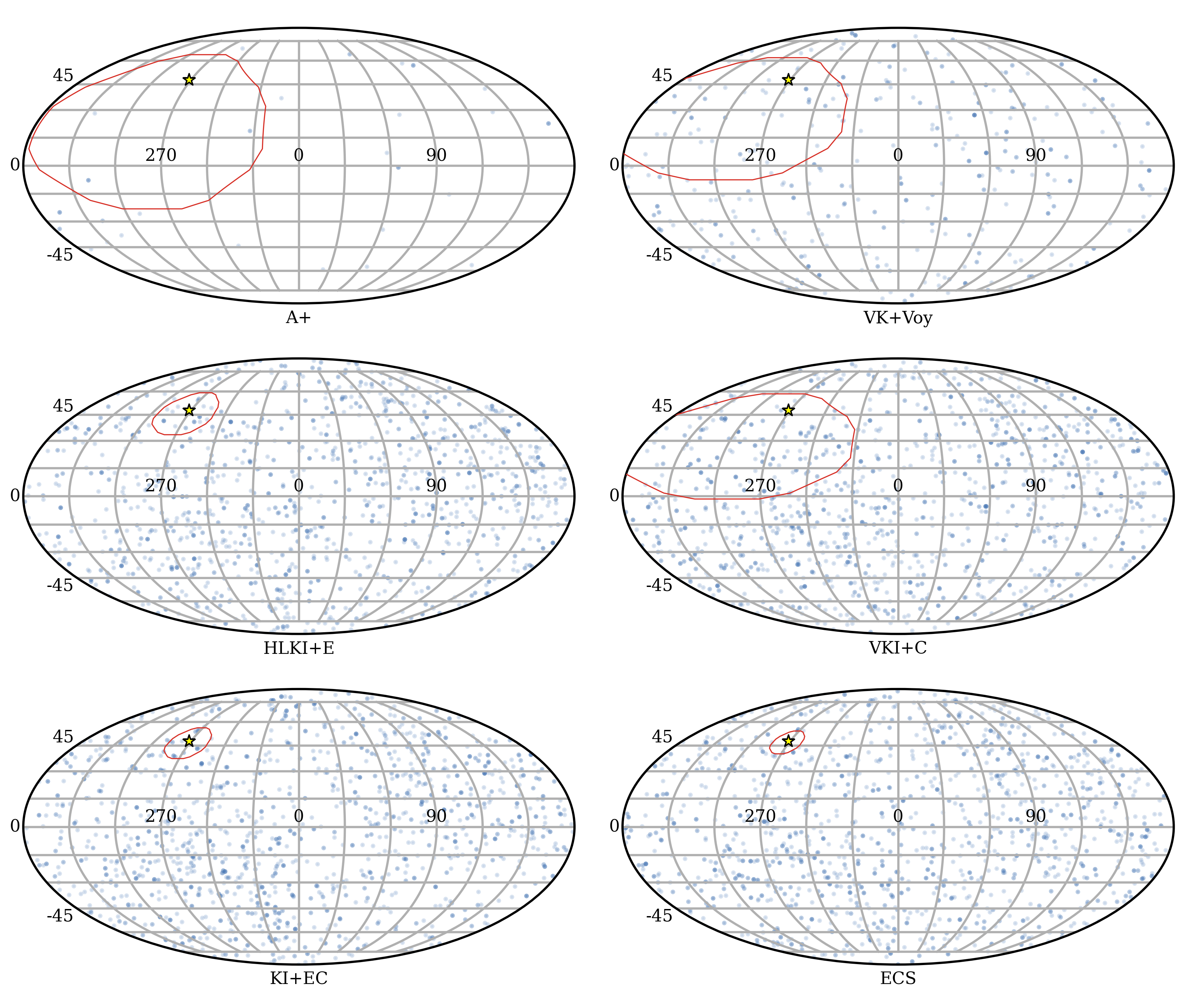}
  \caption{Constraint forecasts on the dipole location for each network. The blue circles in the background are detected injections and the yellow star indicates the true location of the dipole. Each red contour is an ellipse with radii equal to the corresponding network's $1\, \sigma$ projected measurements on the dipole's angular location parameters ($\phi$,\,$\theta$) in galactic coordinates ($l$,$b$); the contour is centered at each network's best-fit location of the dipole. Note that the upgraded CG\texttt{+} and Voyager networks (top) are projected to obtain poor constraints that span nearly half of the sky; the single-XG networks (middle) would improve upon these constraints, however ET would provide much better dipole measurements compared to CE due to breaking the distance--inclination degeneracy as elaborated in the main text. Networks with multiple XG detectors (bottom) would provide superior constraints overall, especially when scaling these single-year results based on multiple years of observations.}
  \label{fig:6net-loc}
\end{figure*}

\begin{table*}[htp!]
    \caption{Projected measurements of the dipole amplitude and location using each network with the corresponding number of bright standard siren detections in each network's simulation set. For reference, the dipole is set to have amplitude $g= 1 \times 10^{-2}$ at galactic coordinates ($l=268\degree$, $b=48\degree$). The last three columns correspond to measurements of each of the three dipole parameters: the amplitude $g$ and location ($l,b$). Observe that the projected constraints all improve with larger simulation sets (with the exception of \cefo as elaborated in the main text).}
    \begin{tabular}{|c|c|c|c|c|}
        \hline
        Network & \multicolumn{1}{c|}{\# of detections} & \multicolumn{1}{c|}{$g \times 10^{-2}$} & \multicolumn{1}{c|}{$l \degree$} & \multicolumn{1}{c|}{$b\degree$}\\
        \hline \hline
        \aplus & 42  & $7\pm 6$ & $259 \pm 77$ & $20 \pm 44$\\ \hline
        \voy & 463 & $2\pm 1$ & $244 \pm 78$ & $27 \pm 35$ \\ \hline
        \et & 1553 & $1.0 \pm 0.2$ & $268 \pm 23$ & $46 \pm 13$ \\ \hline
        \cefo & 1604 & $1.8 \pm 0.9$ & $248 \pm 79$ & $28 \pm 30$ \\ \hline
        \EC & 1760 & $1.0 \pm 0.2$ & $269 \pm 16$ & $47 \pm 9.4$ \\ \hline
        \ECS & 1776  & $1.0 \pm 0.1$ & $268 \pm 12$ & $47 \pm 7.0$ \\ \hline
    \end{tabular}\label{tab:6net-all}
\end{table*}

As such, neither the \aplus nor \voy network is able to place informative constraints on the dipole amplitude nor location. Indeed, as shown in Fig.~\ref{fig:6net-loc}, the $1\sigma$ localization uncertainty contours span a large fraction of the entire sky, indicating that \voy will not be able to accurately measure a dipole even with an order of magnitude larger number of detected injections than \aplus. The only networks that could obtain measurements competitive with existing constraints are those with at least one XG detector, which generally results in $\mathcal{O}(10^{-3})$ constraints on the dipole amplitude and better location constraints than \aplus and \voy.

However, observe in Table~\ref{tab:6net-all} and Fig.~\ref{fig:6net-loc} that not all the XG network constraints are comparable, even though they all have similar numbers of detections. Consider the \cefo and \et networks: they analogously possess a single XG detector (although \et has one more CG\texttt{+} detector than \cefo). \cefo would detect more events than \et, but it would obtain much worse constraints on all dipole parameters. This is a result of \cefo possessing much worse measurement errors for the luminosity distance (shown in Fig.~\ref{fig:aniso_dl}) due to the distinction between ET and CE interferometers discussed in Sec.~\ref{subsec:6netsim}. Thus, the type of XG detector matters greatly due to the importance of both detection counts and luminosity distance measurements in constraining the dipole.

\subsection{Single-network grid simulations}\label{subsec:grid}

\begin{figure*}
    \subfigure[Constraints on the dipole's amplitude $g$.]{
    \includegraphics[width=.75\textwidth]{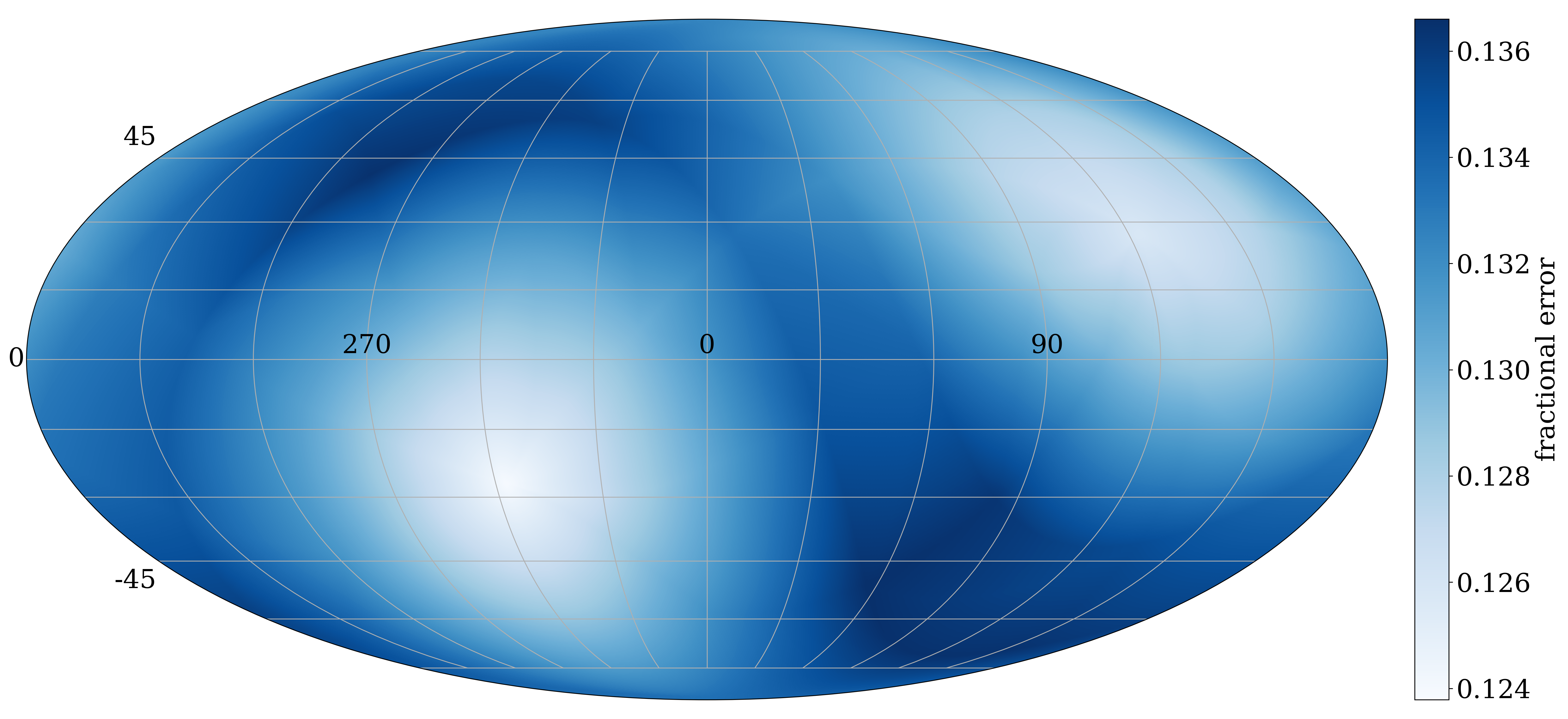}
    \label{subfig:ECS-g}}

    \subfigure[Constraints on the dipole's galactic longitude $l$.]{
    \includegraphics[width=.75\textwidth]{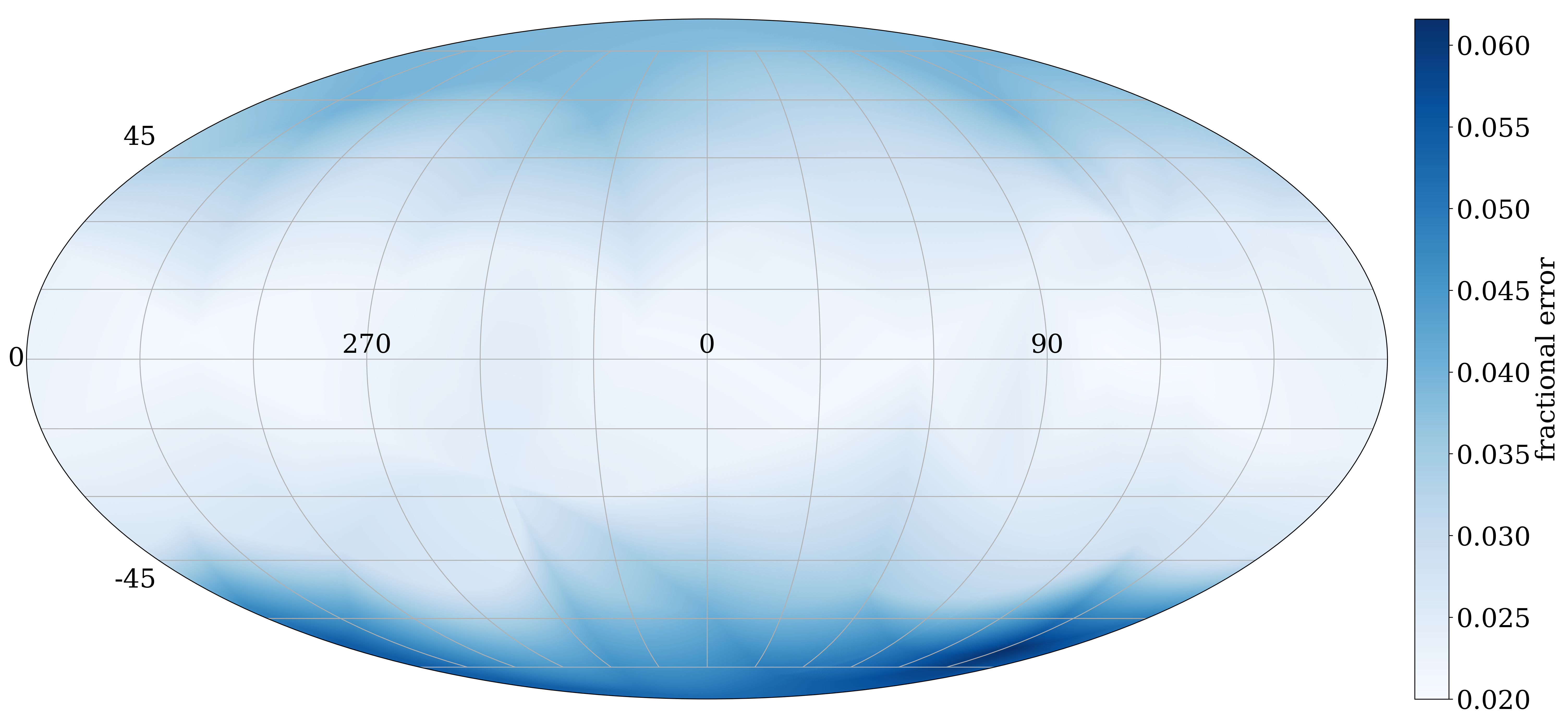}
    \label{subfig:ECS-phi}}

    \subfigure[Constraints on the dipole's galactic latitude $b$.]{
    \includegraphics[width=.75\textwidth]{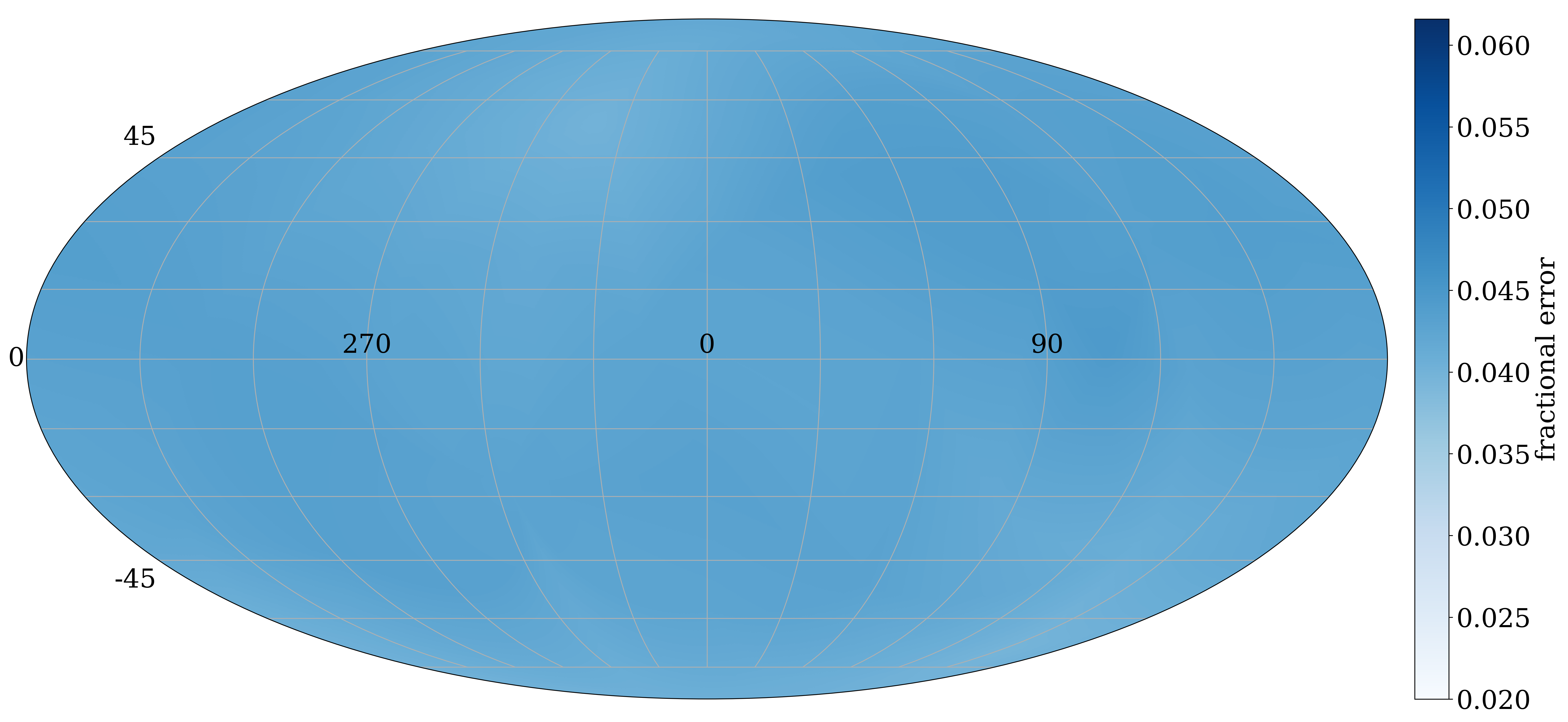}
    \label{subfig:ECS-theta}}

    \caption{Projected constraints for the \ECS network after interpolating across the grid shown in Fig. \ref{fig:lb_grid}, for the amplitude (top), galactic longitude (middle), and galactic latitude (bottom). Note that while there is some variation in the projected constraints across the sky (in particular, for the amplitude and longitude), the absolute range of variation is minimal, $\lesssim 1-4\%$. This indicates that the location of the dipole has only a minor impact on the constraint forecasts.}\label{fig:ECS-grid}
\end{figure*}

We performed the same analysis procedure for each of the 32 simulation sets of the \ECS network. After obtaining the $\chi^2$ constraints on the three model parameters, we interpolated the results across the grid of simulations in galactic coordinates (Fig.~\ref{fig:lb_grid}). The resulting interpolation maps are shown in Fig. \ref{fig:ECS-grid} for the uncertainties on the dipole amplitude $g$ and its galactic coordinate location ($l,b$). Each of these simulations results in constraint forecasts that are comparable in magnitude as those discussed for the \ECS network in Sec.~\ref{subsec:6net}, but with slight variations that depend on the location of the dipole.

In Fig. \ref{subfig:ECS-g}, note that the predicted uncertainty relative to the amplitude of the dipole is generally $\sim 13\%$. The constraint forecast does vary across the sky; this is expected since the dipole's location $\hat{n}$ changes between simulation sets, and thus, $D_L$ will vary in accordance with Eq.~\eqref{eq-dipole}, leading to different estimates of $\sigma_{D_L}$. However, the absolute range of this variation is $\lesssim 1.2\%$ in terms of the fractional error measurement of the amplitude. The constraints are generally best when the dipole is located near the equatorial coordinate poles, but again only by a small margin. This slight variation is consistent with the Poisson noise involved in creating injection sets, which we verified to be $\sim 1.5\%$ by testing injection set generation with different seeds. These results indicate that the location affects the absolute measurement of the amplitude by only $\mathcal{O}(10^{-4}$), so the projected constraints are robust against the dipole's location.

In Fig. \ref{subfig:ECS-phi}, observe that the predicted constraints on the measurements of galactic longitude $l$ relative to the range of $l$ ($360\degree$) is $\sim 2-6\%$, or roughly $6-19\degree$. The dipole location affects these constraints more than those of the amplitude, with the worst constraints occurring around the galactic poles. This is consistent with numerical precision limits when performing astronomical coordinate system transformations in the vicinity of a coordinate singularity. Within \texttt{astropy}, we verified that this effect introduces an error of $\lesssim 3\%$ when performing the transformation from ($\alpha, \delta$) to ($l, b$); thus, the forecasts shown here are upper limits, which nonetheless indicate only minor variation of the constraints as a function of dipole location.

The constraint forecasts for measurements of galactic latitude $b$ are much more consistent across the sky when compared to those for the longitude $l$, as shown in Fig. \ref{subfig:ECS-theta}. Note that the constraints on $b$ relative to the range of $b$ ($180\degree$) are nearly uniform around $\sim 4\%$, with only minor variation on the sky ($\sim 0.5\%$). There are no coordinate singularities involved in the constraints of the latitude, and thus, the issues of numerical precision errors present in the longitude results are not apparent here.

\begin{figure*}
     \subfigure[One year of observations]{\includegraphics[width=0.48\textwidth]{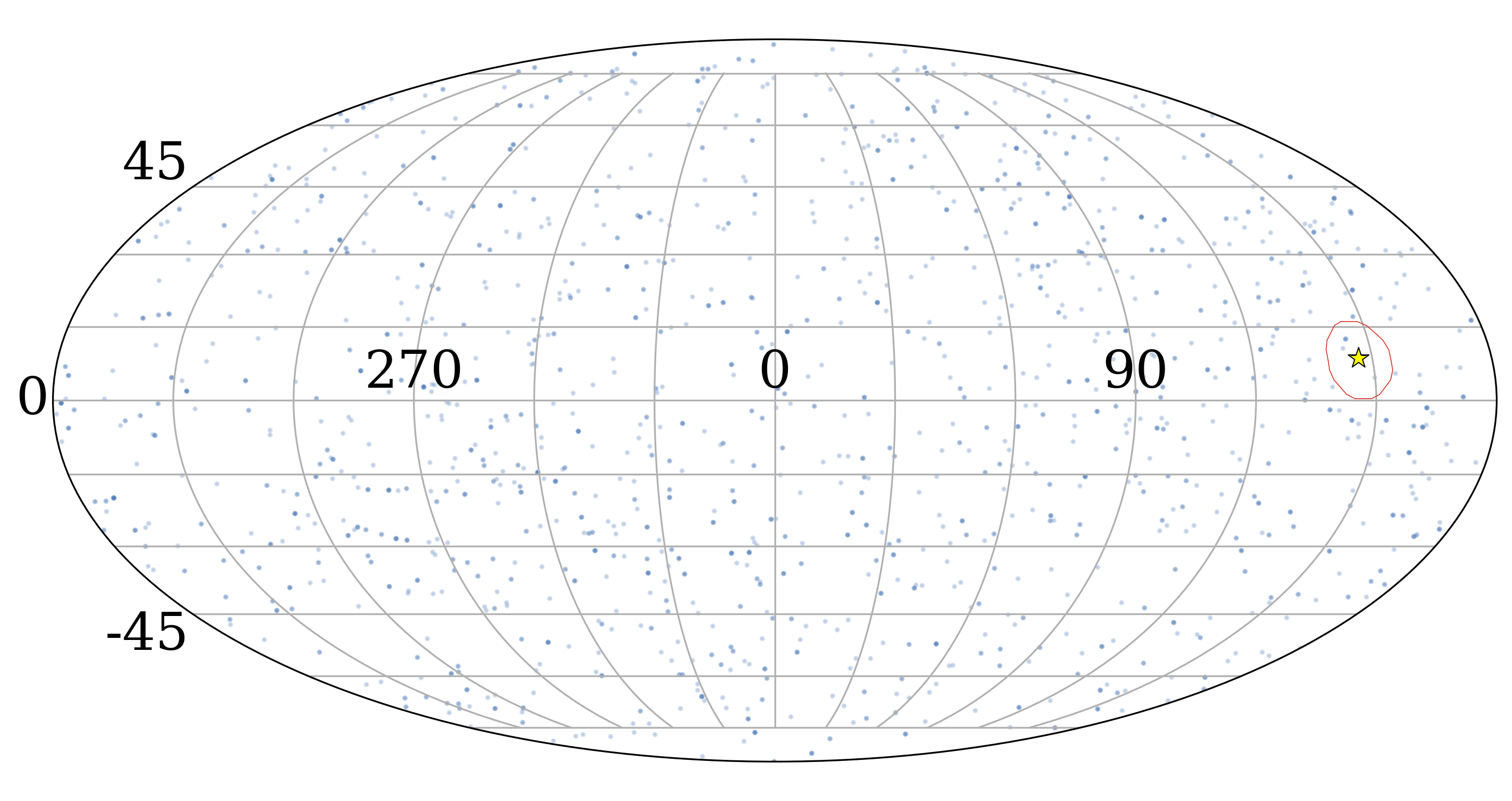}\label{subfig:ECS-1yr}}
     \subfigure[10 years of observations]{\includegraphics[width=0.48\textwidth]{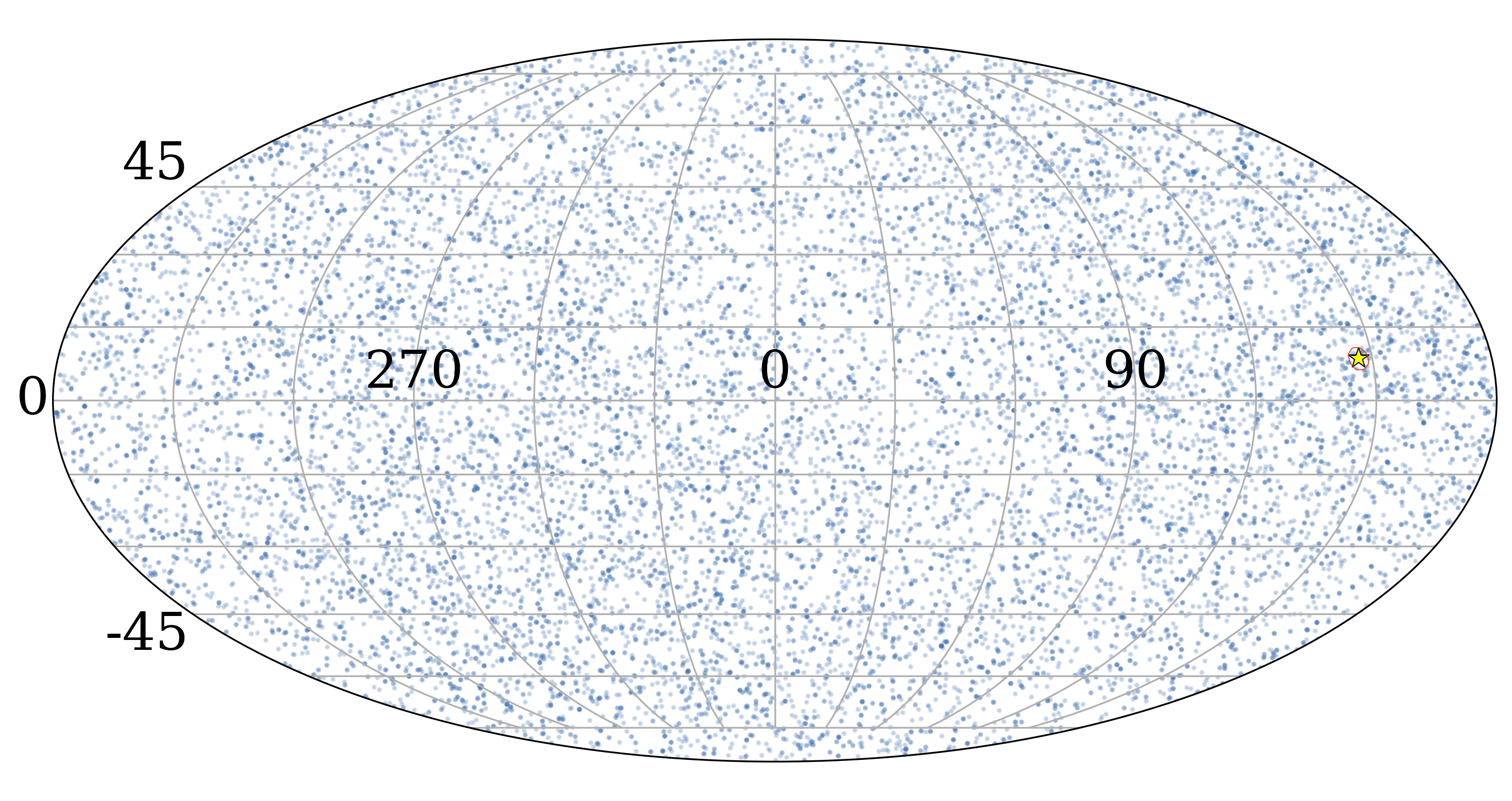}\label{subfig:ECS-10yr}}
     \caption{Constraints on the dipole location for 1-year and 10-years of observations with the \ECS network. The yellow star indicates the true location of the dipole; each red contour is an ellipse with radii equal to the $1\, \sigma$ constraints on the dipole's angular location parameters in galactic ($l,b$). For reference, the dipole is set to have amplitude $g= 1 \times 10^{-2}$ at galactic coordinates ($l=147\degree$, $b=8.6\degree$). Note that the number of detections (blue dots) is dramatically increased with more years of observations, leading to a corresponding improvement in the location constraint.}
  \label{fig:ECS-years-loc}
 \end{figure*}

To investigate the scaling relations of these results, we also present a comparison of our results from 1 year and 10 years of simulated observations for one of the grid points. As shown in Table~\ref{tab:ECS-years}, 10 years of observations yield $10\times$ the number of detected injections $N$, as expected. Observe that the projected constraints improve by a factor of $\sim 3-3.5 \approx \sqrt{10}$ with respect to the 1-year simulation set, as shown for the amplitude in Table~\ref{tab:ECS-years} and the location in Fig.~\ref{subfig:ECS-10yr}. This verifies that the projected constraints improve by $\sim \sqrt{N}$ for the \ECS network. 

\begin{table*}[htp]
    \caption{Forecasted constraints on the dipole amplitude and location with the corresponding number of bright standard siren detections for 1-year and 10-years of observations with the \ECS network. Observe that $\sim 10 \times$ the number of detections are obtained over 10 years, which results in forecasted constraints that are $\sim \sqrt{10}$ smaller, as expected.}
    \begin{tabular}{|c|c|c|c|c|}
        \hline
        Network & \multicolumn{1}{c|}{\# of detections} & \multicolumn{1}{c|}{$g \times 10^{-2}$} & \multicolumn{1}{c|}{$l \degree$} & \multicolumn{1}{c|}{$b\degree$}\\
        \hline \hline
        \ECS (1 year) & 1776  & $1.0 \pm 0.1$ & $147 \pm 7.9$ & $8.2 \pm 7.8$  \\ 
        \hline
        \ECS (10 years) & 17748  & $1.00 \pm 0.04$ & $147 \pm 2.4$ & $8.5 \pm 2.3$ \\
\hline
    \end{tabular}\label{tab:ECS-years}
\end{table*}

\begin{figure*}
    \subfigure[Projected amplitude constraints]{\includegraphics[width=0.48\textwidth]{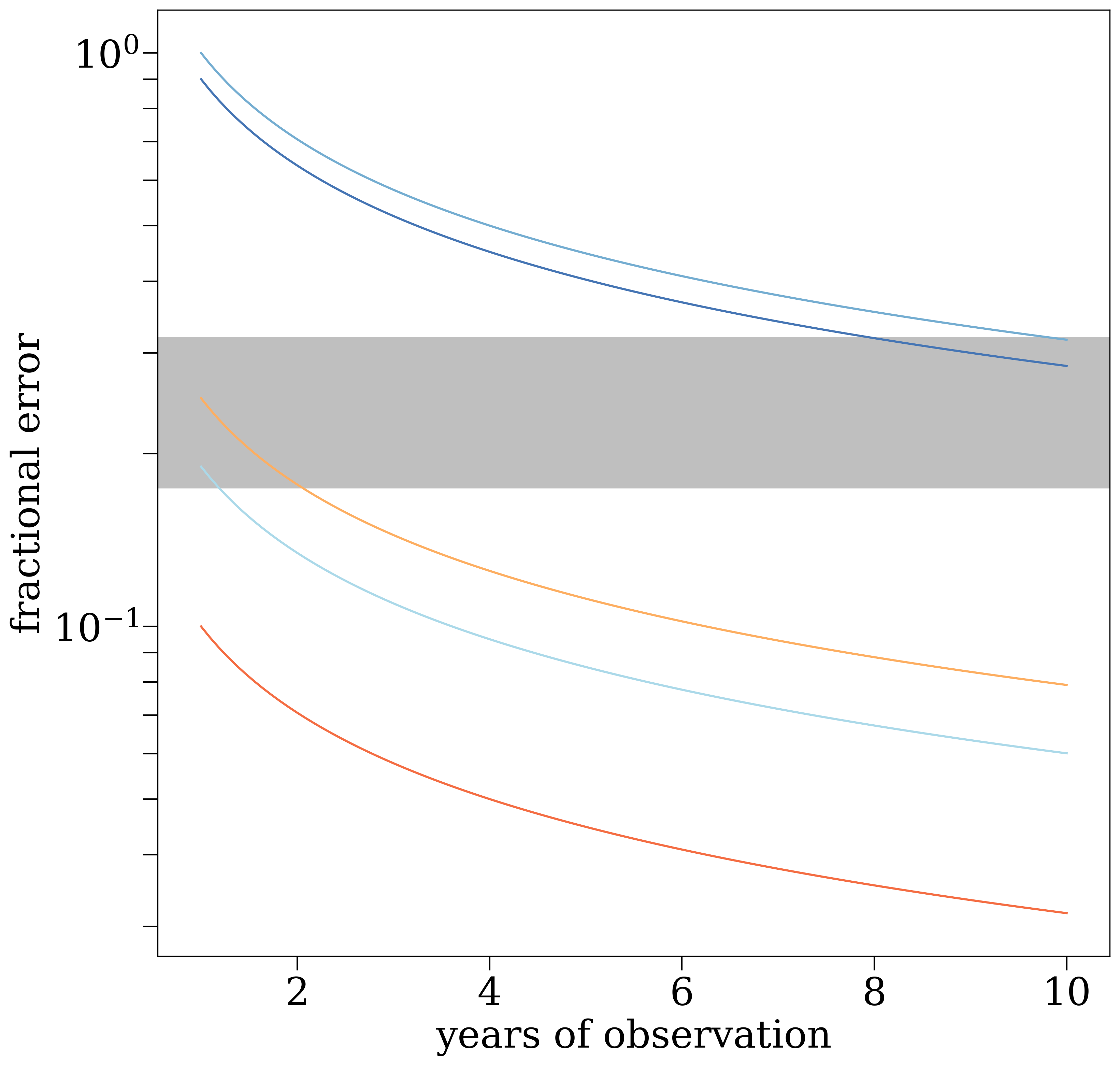}\label{subfig:scale-yr-g}}
    \subfigure[Projected location constraints]{\includegraphics[width=0.48\textwidth]{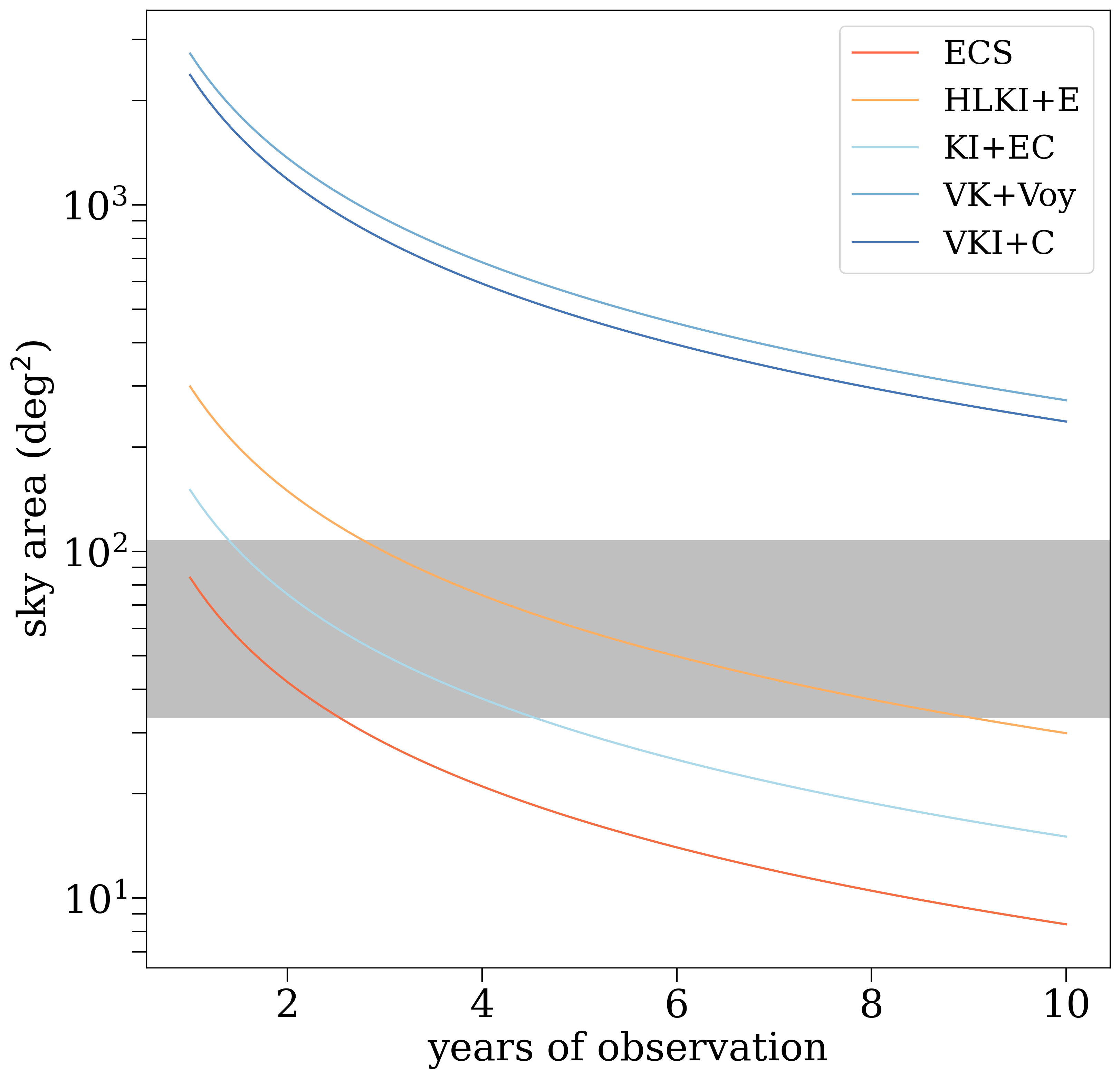}\label{subfig:scale-yr-A}}
    \caption{Dipole constraint forecasts each network as a function of observing time. For comparison, the shaded region corresponds to the range of constraints spanned by current late-universe dipole measurements from Table~\ref{tab:dipoles}. Note that only five networks are shown since \aplus did not exhibit the same scaling behavior. The error on the dipole's amplitude (location) is shown on the left (right). Note that at least one XG detector is required to match or exceed existing constraints. Note also that the dipole amplitude would in general be able to be measured more accurately than the location, relative to existing astrophysical measurements.}\label{fig:scale-yr}
\end{figure*}

The other networks generally follow the same scaling behavior: we verified that their projected constraints improve by at least a factor of $\sim \sqrt{10}$ after 10 years of observations. Thus, we also present the scaling of the constraints as a function of years of observation (Fig.~\ref{fig:scale-yr}). However, the constraints from the \aplus network did not follow the scaling relation: even after 10 years of observations, its dipole constraints were still $\sim 100\%$ for the amplitude and $\mathcal{O}(10^3) \, \mathrm{deg}^2$ for the direction, and thus we exclude it from the scaling results.

Fig.~\ref{fig:scale-yr} provides a comparison between our constraint forecasts and existing measurements. The shaded regions show the range of constraints spanned by two current late-universe dipole measurements from Table~\ref{tab:dipoles} (tightest constraints from \cite{bengalyProbingCosmologicalPrinciple2018} and widest constraints from \cite{singalPeculiarMotionSolar2022}). This shows that GW networks would generally be able to measure the amplitude of a dipole more accurately than the location, at least relative to existing dipole measurements. Note that a network with one ET would be sufficient to obtain constraints superior to existing measurements of a dipole, but only after multiple years of observations ($\sim 2$ years for the amplitude and $\sim 9$ years for the location); multiple XG detectors would obtain superior constraints much more quickly.

Additionally, Fig.~\ref{fig:scale-yr} provides a detailed picture of the capabilities between future networks. For example, for both the amplitude and location, one year of observing with the \et network is expected to produce constraints comparable to \textit{10} years of observations with the \voy or \cefo networks (even with $\lesssim 1/3$ the number of cumulative detections). Conversely, the 10-year amplitude constraints from the \et network would be nearly matched after just one year of observations from the \ECS network. 

\section{Discussion and Conclusions} \label{sec:conclusion}
In this work, we have investigated the prospects of using networks of CG and XG detectors to measure a general dipole anisotropy in the luminosity distance via BNS bright standard sirens. In doing so, we have addressed several important considerations required to forecast dipole measurements. Firstly, we created more realistic simulations of GW signals by implementing a state-of-the-art GW model for the signal injections. Secondly, we made the GW detector scenarios more realistic by considering the detectors to be physically located across the surface of Earth, which accounts for their anisotropic sky sensitivity as well as the GW signals' phase differences between different detectors. We also allowed for Earth's rotation, thus capturing the dynamic time-evolution of the detectors' sensitivities. Finally, we took a model-agnostic, phenomenological approach to measuring an anisotropy by modeling a generic dipole in luminosity distance and considering a scenario wherein the location of the dipole varies across the sky. We addressed these considerations by utilizing \texttt{gwbench} to perform simulation campaigns for a variety of detector networks and dipole locations.

We found that the types and locations of detectors matter substantially in the ability to measure a dipole after a year of observations. Upgraded CG detectors alone, which would find $\mathcal{O}(10^{1-2})$ bright standard sirens annually, are not able to informatively constrain a dipole; competitive constraints would not possible until at least one XG detector is available, allowing for $\mathcal{O}(10^3)$ detections. We found that for a dipole of amplitude $0.01$ coincident with the CMB dipole, a network with a single XG detector observing for one year would at best measure the dipole amplitude comparably to previous studies that use electromagnetic probes. In particular, when comparing to previous studies referenced in lines 4-6 of Table~\ref{tab:dipoles}, the \et network would obtain competitive amplitude constraints ($\sim 20\%$ versus $\sim 17-33\%$) but inferior direction constraints ($\sim 300 \,\mathrm{deg}^2$ versus $\sim 33-108 \,\mathrm{deg}^2$) after one year of observations. When multiple XG detectors are considered, the constraints are expected to be comparable or superior to some existing measurements. For example, we found that the \ECS network could measure the dipole amplitude at $\sim 10\%$ (compared to $\sim 17-33\%$) and the dipole direction to $\sim 84 \, \mathrm{deg}^2$ (compared to $\sim 33-108 \,\mathrm{deg}^2$). 

Moreover, we demonstrated that the type of XG detector is vital: while a network with a single CE would obtain $\mathcal{O}(10^3)$ detections, it would obtain worse constraints than a network with ET that detects fewer events. This is ultimately because of the differing accuracy at which each network can measure luminosity distances. ET is planned to contain multiple interferometers that make it capable of measuring GW polarizations---and hence binary inclinations and luminosity distances---very accurately, whereas doing the same with CE would require a separate interferometer of comparable sensitivity in the network. This effect plays a large role in our results given that we consider a generic dipole in the luminosity distance, but this may not apply to other dipole models. A dipole in the number counts~\cite{mastrogiovanniDetectionEstimationCosmic2023}, for example, would be less sensitive to which XG detector is utilized in the network since any XG detector would offer at least 1-2 orders of magnitude more detections than non-XG networks. Contrarily, a dipole that involves both number counts and distances of events~\cite{grimmCombiningChirpMass2023} may be only slightly impacted by this effect.

We also demonstrated that the forecasted constraints scale with the square root of number of detections $N$, and thus the constraints provided in this paper can be adjusted to fit scenarios that we did not examine here. For example, in the context of multi-messenger astronomy in the XG era~\cite{chenProgramMultiMessengerStandard2021}, there may be EM follow-up criteria that differ from what we considered in the present work (e.g., a more stringent cut on the redshift or sky localization of an event), but in this case, our projected constraints can simply be scaled accordingly based on $N$. The constraint forecasts can be similarly scaled to obtain estimates for multiple years of observing: three years of observations would yield $3\times$ more detections and thus the constraints would simply be smaller by a factor of $\sqrt{3}$.

We applied this scaling to our results for the XG networks to show that constraints superior to existing methods would be obtained after multiple years of observations for the \et, \EC, and \ECS networks. For example, after just three years of observing, \ECS would obtain constraints on the amplitude to $\sim 6\%$ and direction to $28 \, \mathrm{deg}^2$, a marked improvement upon existing measurements obtained by dedicated, multi-year surveys of supernovae~\cite{singalPeculiarMotionSolar2022,horstmannInferenceCosmicRestframe2022,betouleImprovedCosmologicalConstraints2014} and galaxies~\cite{bengalyProbingCosmologicalPrinciple2018,condonNRAOVLASky1998,intemaGMRT150MHz2017}. But, as a general caveat, note that the nature of the dipoles are not exactly the same form (e.g., the tightest constraints we compare with here come from a dipole measured by the distortion of radio galaxy counts~\cite{bengalyProbingCosmologicalPrinciple2018}, not a strict dipole in the luminosity distance), making the comparison imperfect. Nonetheless, the dipole constraints we compared with are not fully explained by kinematics and thus may possess a partial intrinsic component, making them more adequate subjects of comparison than, e.g., the CMB kinematic dipole~\cite{planckcollaborationPlanck2018Results2020a}.

We also found that the dipole's location would play only a minor role in its measurements. When considering a network of three XG detectors and a variable dipole location, we found that the absolute variation in the forecasted amplitude constraints is only $\sim 1.2\%$ across the sky, and the corresponding variation for the location is $\sim 2-6\%$. These results indicate that the location contributes an absolute error on the amplitude measurement of only $\mathcal{O}(10^{-4})$ for a $g=0.01$ dipole amplitude, even when considering the anisotropic sky sensitivities of the detector networks. This result is reassuring in light of the concerns about neglecting detector sky direction dependence noted in~\cite{mastrogiovanniDetectionEstimationCosmic2023}, which anticipated effects of absolute order $\mathcal{O}(10^{-3})$ on measurements of the dipole amplitude.

Our results complement existing studies that consider GWs as probes of cosmic dipoles~\cite{stiskalekAreStellarmassBinary2020,essickIsotropyMeasurementGravitational2023,kashyapDipoleAnisotropyGravitational2023,grimmCombiningChirpMass2023,caiProbingCosmicAnisotropy2018,caiProbingCosmicAnisotropy2019}. While we considered several ideas that were previously unaddressed in the literature, there are other questions and applications that remain open. For example, we considered BNS mergers as bright standard sirens, while similar investigations could be carried out using BBHs as dark standard sirens~\cite{schutzDeterminingHubbleConstant1986}; such a consideration would greatly improve the detection number counts due to the $\mathcal{O}(10^5)$ annual BBH detections expected for XG detectors~\cite{evansHorizonStudyCosmic2021,maggioreScienceCaseEinstein2020}, but the imperfect 2D sky localization of these events may limit their usage for dipole measurements. We also used only a single dipole amplitude in our simulations; while this fiducial value was selected based on existing dipole amplitudes, the usage of a range of dipole values would provide a more comprehensive set of dipole measurement forecasts. Additionally, in our studies of varying dipole location, we examined only the best-case network (\ECS), whose results may not be exactly representative of less-sensitive networks with one or two XG detectors.

Our work highlights another important contribution that XG detectors would have for addressing open questions in cosmology. Beyond demonstrating a promising application of the exquisite sensitivity of XG detectors, our work shows that the type of XG detector(s) that are constructed will have a large impact on certain scientific outcomes (in this case, measuring a luminosity distance dipole). This underscores the importance of carefully considering the scientific motivations underlying different network designs during the planning of ground-based XG detectors.


\acknowledgements
B.C. thanks Gilbert Holder and Dragan Huterer for illuminating discussions on cosmological anisotropies.
B.C.~acknowledges that this material is based upon work supported by the National Science Foundation (NSF) Graduate Research Fellowship Program under Grant No. DGE 21-46756. N.Y. is supported from the Simons Foundation through Award No. 896696 and the NSF Grant No. PHY-2207650. B.S.S. is supported through NSF grants AST-2307147, PHY-2207638, PHY-2308886, and PHY-2309064. Computing resources for the creation and analysis of these simulations were provided by the Gwave supercomputer\footnote{\href{https://computing.docs.ligo.org/guide/computing-centres/psu/}{https://computing.docs.ligo.org/guide/computing-centres/psu/}}, International Gravitational-Wave Observatory Network~\cite{bagnasco2024ligo}, and the Pennsylvania State University Institute for Computational and Data Sciences\footnote{\href{https://www.icds.psu.edu/}{https://www.icds.psu.edu/}}.

\textbf{Software:} \texttt{astropy}~\cite{astropy2022}, \texttt{h5py}~\cite{collette2013python}, \texttt{matplotlib}~\cite{hunter2007matplotlib}, \texttt{numpy}~\cite{harris2020numpy}, \texttt{scipy}~\cite{virtanen2020scipy}.
\pagebreak

\bibliography{bibliography}

\end{document}